\documentclass[12pt]{article}
\usepackage{epsfig,amsmath,amssymb,amsfonts,amstext,amsthm}
\usepackage{latexsym,graphics,epsf,epsfig,cite}

\newcommand{\be}{\begin{eqnarray}}
\newcommand{\ee}{\end{eqnarray}}

\newcommand{\qq}{\vspace*{-0.1in}}

\newcommand{\vx}{{{\mathbf{x}}}}
\newcommand{\vy}{{{\mathbf{y}}}}
\newcommand{\vz}{{{\mathbf{z}}}}
\newcommand{\vw}{{{\mathbf{w}}}}
\newcommand{\vu}{{{\mathbf{u}}}}
\newcommand{\va}{{\mathbf{A}}}
\newcommand{\vi}{{\mathbf{I}}}
\newcommand{\vd}{{\mathbf{D}}}
\newcommand{\vk}{{\mathbf{K}}}
\newcommand{\vl}{{\mathbf{\Lambda}}}
\newcommand{\vq}{{\mathbf{Q}}}

\newcommand{\vo}{{\mathbf{0}}}
\newcommand{\diag}{{\text{diag}}}
\newcommand{\cov}{{\text{Cov}}}
\newcommand{\meane}{{\mathbb{E}}}
\newcommand{\listl}{{l=1,\;\dots,\;L}}

\newcommand{\mB}{\bf B}

\newtheorem{lemma}{\textbf{\textsl{Lemma}}}
\newtheorem{theorem}{\textbf{\textsl{Theorem}}}
\newtheorem{proposition}{\textbf{\textsl{Proposition}}}

\newcommand{\df}{\stackrel{{\rm def}}{=}}

\begin{document}

\vspace*{-2cm}
\begin{center}
\em{Submitted to the IEEE Transactions on Information Theory, Oct.
2005}
\end{center}

\begin{center}
  \baselineskip 1.3ex {\Large \bf Vector Gaussian Multiple Description with Individual and Central
Receivers\footnote{This research was sponsored in part by
  NSF CCR-0325924 and a Vodafone US Foundation Fellowship.}} \\

\vspace{0.15in} Hua Wang and Pramod Viswanath \footnote{The
authors are with the Department of Electrical and Computer
Engineering and the Coordinated Science Laboratory, University of
Illinois at Urbana-Champaign, Urbana IL~~61801; e-mail: {\tt
 \{huawang,pramodv\}@uiuc.edu}}
\end{center}

\begin{abstract}
$L$ multiple descriptions of a vector Gaussian source for
individual and central receivers are investigated. The sum rate of
the descriptions with covariance distortion measure constraints,
in a positive semidefinite ordering, is exactly characterized. For
two descriptions, the entire rate region is characterized. Jointly
Gaussian descriptions are optimal in achieving the limiting rates.
The key component of the solution is a novel information-theoretic
inequality that is used to lower bound the achievable multiple
description rates.

\end{abstract}
\vspace{-0.2in} \baselineskip\normalbaselineskip
\qq
\section{Introduction} \label{sec:intro}
\qq

In the multiple description problem, an information source is
encoded into $L$ packets and these packets are sent through
parallel communication channels. There are several receivers, each
of which can  receive a subset of the packets and needs to
reconstruct the information source based on the received packets.
In the most general case, there are $2^L -1$ receivers and the
packets received in each receiver correspond to one of $2^L -1$
subsets of $\{1, \; \dots, \; L\}$. A long standing open problem
in the literature \cite{Ozarow80,ElGamal82,Ahlswede85,Zhang87,Zamir99,FWFu02,Venkat03,Pradhan04,HYFeng05,Puri05}  is to characterize the information-theoretic
rate region subject to the specified  distortion constraints.
Practical multiple description codes have  been discussed
 in
\cite{Vaishampayan93,Vaishampayan94,Vaishampayan98,Vaishampayan01,Goyal01,Diggavi02,Goyal02,CTian04} and
recent work \cite{Ishwar03,JChen05} has considered the  multiple description problem in the context of  the distributed source coding scenario.
Optimal descriptions of even the Gaussian source with quadratic
distortion measures have not been fully characterized. In the special case
of two descriptions of  a scalar Gaussian source with quadratic
distortion measures, however, the  entire rate region has been
characterized in \cite{Ozarow80}.

\begin{figure}
\begin{center}
\scalebox{1.2}{ \input{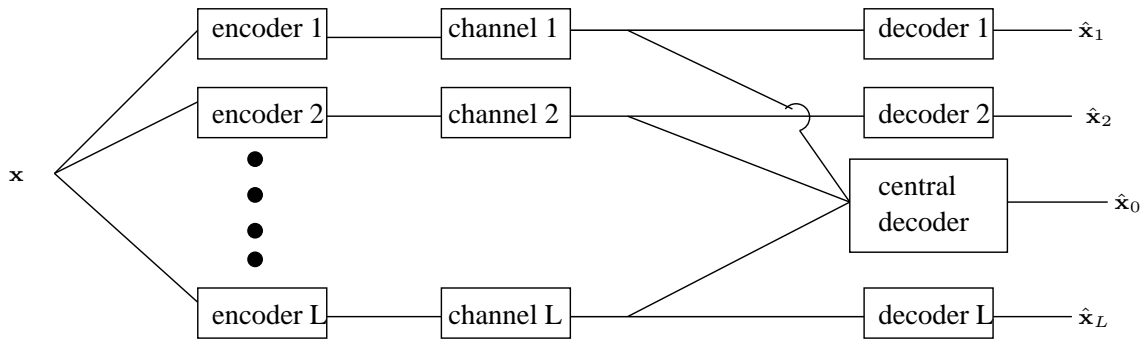} }
\caption{MD problem with only individual reconstructions and
central reconstruction} \label{fig:md}
\end{center}
\end{figure}

Our focus  is on $L$
descriptions of a memoryless {\em vector} Gaussian source forwhere
 $L$ individual  and a single common receiver (cf.
Figure~\ref{fig:md}). Each receiver needs to reconstruct the
original source such that the empirical covariance matrix of the
difference is less than, in the sense of a  positive semidefinite
ordering, a ``distortion'' matrix. In this setting,  the
  symmetric rate multiple description problem of a scalar Gaussian
source with symmetric distortion constraints has been characterized
in \cite{Venkat03,Pradhan04,Puri05}, but  a complete understanding of
all other  rate-distortion settings is open.

Our main result is an {\em exact} characterization of the sum rate for
any specified $L+1$  distortion matrix
constraints. With  $L=2$, we  characterize the entire
rate region. Our contribution is  two fold:
\begin{itemize}
\item First, we derive a novel information-theoretic inequality that
 provides a  lower bound to the sum of  the
description rates. The key step is to avoid using the  entropy power
inequality, which was a central part of the proof of two descriptions
 of the scalar Gaussian source in \cite{Ozarow80}: the vector  entropy
power inequality is tight only  with a  certain
covariance alignment condition, which arbitrary distortion matrix
 requirements do not necessarily allow.
\item Second, we show that  jointly Gaussian  descriptions actually
achieve the lower bound  not by resorting to a direct calculation
and comparison, which appears to be difficult for $L > 2$, but
instead by arguing
the equivalence of certain optimization problems. 
\end{itemize}

\begin{figure}
\begin{center}
\scalebox{1.2}{ \input{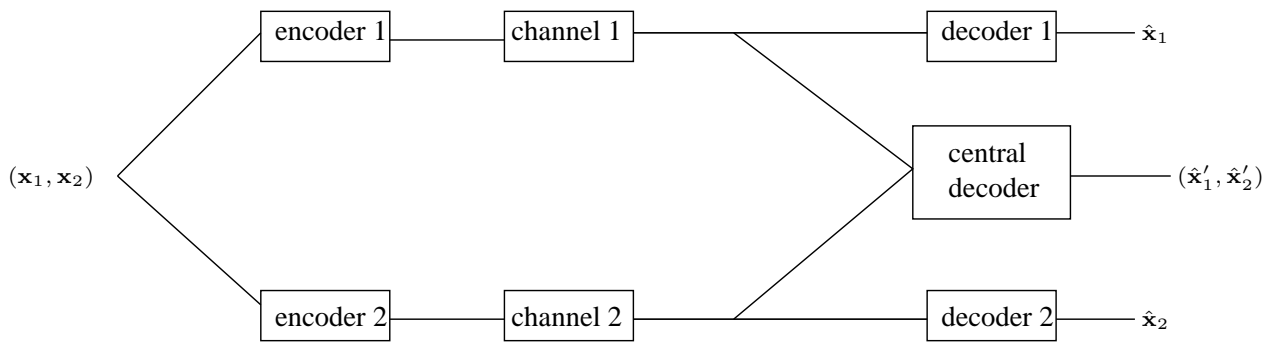} }
\caption{Multiple Descriptions with separate distortion
constraints.} \label{fig:md_special}
\end{center}
\end{figure}

Consider another two description problem of a pair of jointly Gaussian
memoryless sources as depicted in
Figure~\ref{fig:md_special}.   There are two encoders that describe this
source to three receivers: receiver $i$ gets the description of
encoder $i$, with $i=1,2$ and the third receiver receives both the
descriptions. Suppose receiver $i$ is interested in reconstructing
the $i$th marginal of the jointly Gaussian source, with $i=1,2$.
The third receiver is interested in reconstructing the entire vector
source.  This description problem is closely related to the vector
Gaussian description problem that is the main focus of this paper.
We exploit this connection and characterize the rate region where
the reconstructions have a constraint on the covariance of error
at each of the receivers (in the sense of a positive semidefinite
order).

We have organized the results in this  paper as follows. In Section~\ref{sec2} we give a formal
description of the problem and  summarize our main result. The
derivation of a  lower bound is in Section~\ref{sec3}. In Section~\ref{sec4} we
provide an upper bound and provide conditions for the achievable sum
rate to meet the lower bound. We see  in Section~\ref{sec5}  that the
 conditions are indeed satisfied in the special case of a  scalar
Gaussian source. The solution in the case of the more complicated  vector Gaussian
source is in Section~\ref{sec6}. The solution to  the multiple
description problem depicted in Figure~\ref{fig:md_special} is the
topic of Section~\ref{sec71}. Finally,
while the characterization of the rate region of  general  multiple
descriptions of the Gaussian source (with each receiver having access
to some subset of the descriptions)  is
still open,  we can use the insights derived via our sum rate
characterization to solve this problem for a  nontrivial set of
covariance distortion constraints; this is done in Section~\ref{sec72}.

A note about the notation in this paper: we use lower case
letters for scalars, lower case and bold face for vectors, upper
case and bold face for matrices. The superscript $t$ denotes
matrix transpose. We use $\vi$ and  $\mathbf{0}$ to denote the
identity matrix and the
 all zero matrix respectively, and $\diag\{p_1, \dots,
p_n\}$ to denote a diagonal matrix with the diagonal entries equal
to $p_1, \dots, p_n$. The partial order $ \succ$ ($\succcurlyeq$)
denotes positive definite (semidefinite) ordering:  $\va \succ
\mB$ ($\va \succcurlyeq \mB$) means that $\va -\mB$ is a positive
definite (semidefinite) matrix. We write $\mathcal{N}(\mu, \vq)$
to denote a Gaussian random vector with mean $\mu$ and covariance
$\vq$. All logarithms in this paper are to the natural base.

\section{Problem Setting and Main Results}
\label{sec2}

\subsection{Problem Setting}

The information source $\{\vx[m]\}$ is an i.i.d.\ random process
with the marginal distribution $\mathcal{N}(0, \mathbf{K}_x)$,
i.e., a collection of i.i.d.\  Gaussian random vectors. Denoting
the  dimension of $\{\vx[m]\}$ by $N$, we suppose that $\vk_x$ is
an $N \times N$ positive definite matrix. There are $L$ encoding
functions at the source, encoder $l$ encodes a  source sequence,
of length $n$, $\vx^n = (\vx[1],\;\dots , \; \vx[n])^t$ to a
source code $C^{(n)}_l = f_l^{(n)}(\vx^n)$, for $l=1\ldots L$.
 This code $C^{(n)}_l$
is sent through $l$th communication channel at the rate $R_l =
\frac{1}{n}\log|C^{(n)}_l|$. There are $L$ individual receivers and
one central receiver.

For $l=1, \; \dots \; L$, the $l$th individual  receiver uses its
information (the output of the $l$th channel) to generate an
estimate $\hat{\vx}_l^n$
$=g_l^{(n)}\left(f_l^{(n)}(\vx^n)\right)$ of the source sequence
$\vx^n$. The central  receiver uses  the output of all the $L$
channels to generate an estimate $\hat{\vx}_0^n$
of the source sequence $\vx^n$. Since we are interested in
covariance constraints, the decoder maps can be restricted to be
the minimal mean square error (MMSE) estimate of the source
sequence based on the received codewords. So,
\begin{equation}
\begin{split}
\hat{\vx}_l^{n} & = \meane\left[\vx^n|f_l^{(n)}(\vx^n)\right], \quad l=1,\;\dots,\;L \\
\hat{\vx}_0^{n} & = \meane\left[\vx^n|f_1^{(n)}(\vx^n), \; \dots,
\; f_L^{(n)}(\vx^n)\right].
\end{split}
\end{equation}
Suppose  the reconstructed sequences satisfy the covariance
constraints
\begin{equation}
\begin{split}
\frac{1}{n}\sum\limits_{m=1}^n
\meane\Big[(\vx[m]-\hat{\vx}_l[m])^t(\vx[m]-\hat{\vx}_l[m])\Big] &
\preccurlyeq \mathbf{D}_l,
\quad l=1,\;\dots,\;L, \\
\frac{1}{n}\sum\limits_{m=1}^n
\meane\Big[(\vx[m]-\hat{\vx}_0[m])^t(\vx[m]-\hat{\vx}_0[m])\Big] &
\preccurlyeq \mathbf{D}_0,
\end{split}
\end{equation}
then we say that multiple descriptions with distortion constraints
$(\vd_1,\;\dots,\;\vd_L,\;\vd_0)$  are achievable  at the rate
tuple $(R_1, \; \dots, \; R_L)$.


The closure of the set of all achievable rate tuples is called the
rate region and is denoted by $\mathcal{R}_*(\vk_x, \;
\vd_1,\;\dots,\;\vd_L,\;\vd_0)$.
Throughout this paper, we suppose that  $\mathbf{0} \prec \vd_0 \prec
\vd_l \prec \vk_x,\; \forall
l=1,\dots,L$.\footnote{That $\vd_0 \preccurlyeq \vd_l$, is without
  loss of generality is seen by applying the data processing
  inequality for mmse estimation errors; having more access to
  information can only reduce the covariance of the error in a
  positive semidefinite sense. Similarly, $\vk_x \preccurlyeq \vd_0$
  is also not interesting; here we simplify this condition and take
  $\vd_0 \prec  \vk_x$.}

\subsection{Sum Rate}\label{sec:sumrate}

Our main result is the precise characterization of the sum rate of
multiple descriptions for  individual and central receivers.
\begin{theorem}
For distortion constraints
$(\vd_1,\;\dots,\;\vd_L,\;\vd_0)$, the sum rate 
is
\begin{equation}\label{eq:sumrate}
\displaystyle 
\sup_{\vk_z \succ
\mathbf{0}}\quad
\frac{1}{2}\log\left (\frac{|\vk_x||\vk_x+\mathbf{K}_z|^{(L-1)}|\mathbf{D}_0+\mathbf{K}_z|}
{|\mathbf{D}_0|\prod\limits_{l=1}^L|\mathbf{D}_l+\mathbf{K}_z|}\right ).
\end{equation}
\end{theorem}
This sum rate is achieved by a {\em jointly Gaussian random
multiple description scheme}: let $\vw_1,\; \cdots, \; \vw_L$ be
zero mean jointly Gaussian random vectors independent of $\vx$,
with the   positive definite covariance matricex $(\vw_1,\;
\cdots, \; \vw_L)$ denoted by $\vk_w$. Defining
\[
\vu_l = \vx + \vw_l, \quad l=1, \; \dots, \; L,
\]
we consider  $\mathbf{K}_w$ such that
\begin{equation}\label{eq:covariance}
\begin{split}
\cov[\vx|\vu_l] \overset{\text{def}}{=} &
\meane\Big[(\vx-\meane[\vx|\vu_l])^t(\vx-\meane[\vx|\vu_l])\Big]
 \preccurlyeq \mathbf{D}_l,
\quad l=1,\;\dots,\;L, \\
\cov[\vx|\vu_1,\;\dots,\;\vu_L] \overset{\text{def}}{=} &
\meane\Big[(\vx-\meane[\vx|\vu_1,\;\dots,\;\vu_L])^t(\vx-\meane[\vx|\vu_1,\;\dots,\;\vu_L])\Big]
  \preccurlyeq \mathbf{D}_0.
\end{split}
\end{equation}
To construct the code book for the $l$th description, draw
$e^{nR_l}$ $\vu^n_l$ vectors randomly according to the marginal of
$\vu_l$. The encoders observe the source sequence $\vx^n$,
look for codewords $(\vu^n_1, \; \dots, \; \vu^n_L)$ that are
jointly typical with $\vx^n$ and send the index of the resulting $\vu^n_l$
 through the $l$th channel, respectively. The $l$th
individual
 receiver uses  this index and generates a reproduction sequence
$\meane[\vx^n|\vu^n_l]$ for $l=1\ldots L$, the central  receiver uses all the
$L$ indices to generate a reproduction sequence
$\meane[\vx^n|\vu^n_1,\;\dots,\;\vu^n_L]$. For every $\vk_w$
satisfying \eqref{eq:covariance}, the rate tuple
$(R_1,\;\dots,\;R_L)$ satisfying
\begin{equation}\label{eq:innerboundinu}
\sum\limits_{l \in S}R_l \ge \sum\limits_{l \in S}h(\vu_l) -
h(\vu_l, l \in S|\vx) = \frac{1}{2}\log\frac{\prod\limits_{l \in
S}|\vk_x+\vk_{w_l}|}{|\vk_{w_S}|}, \quad \forall S \subseteq \{1,
\; \dots, \; L\}
\end{equation}
is achievable by using this coding scheme, where $\vk_{w_S}$ is
the covariance matrix for all $\vw_l, l \in S$, and $\vk_{w_l} =
\meane[\vw_l^t\vw_l]$. In particular, the achievable sum rate is
\begin{equation}\label{eq:achievablesumrate}
\frac{1}{2}\log\frac{\prod\limits_{l=1
}^L|\vk_x+\vk_{w_l}|}{|\vk_{w}|}.
\end{equation}


We denote this ensemble of descriptions, throughout this paper, as
the jointly Gaussian description scheme and the time sharing
between them as the jointly Gaussian description strategy. We show
that jointly Gaussian description schemes are optimal in achieving
the sum rate \eqref{eq:sumrate}.



\subsection{Rate Region for Two Description Problem}

\begin{figure}[h]
\begin{center}
\scalebox{0.6}{ \input{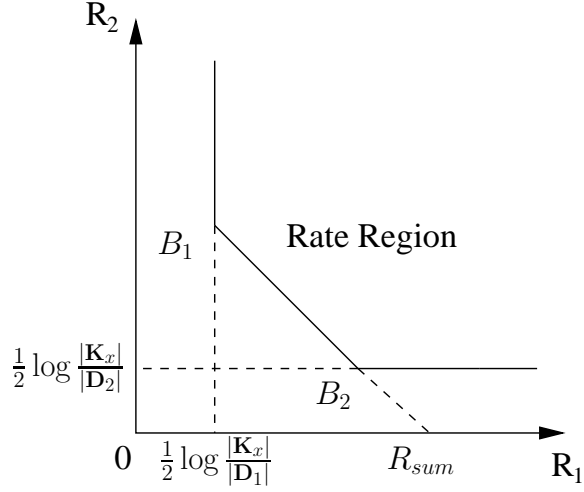} }
          \caption{Rate region for two description problem}
          \label{fig:region}
          \end{center}
\end{figure}

For two descriptions, we can characterize the entire
 rate region.
\begin{theorem}
Given distortion constraints $(\vd_1,\;\vd_2,\;\vd_0)$, the rate
region for the two description problem for an i.i.d.\
$\mathcal{N}(0,\vk_x)$ vector Gaussian source is
\begin{equation}\label{eq:innerboundl2}
\mathcal{R}_*(\vk_x,\;\vd_1,\;\vd_2,\;\vd_0) = \left\{
\begin{array}{l}
(R_1,\;R_2): \\
\displaystyle R_l \ge \frac{1}{2}\log\frac{|\vk_x|}{|\vd_l|}, \quad l=1,\;2 \\
\displaystyle R_1+R_2 \ge \sup_{\vk_z \succ \mathbf{0}}
\frac{1}{2}\log\frac{|\vk_x||\vk_x+\mathbf{K}_z||\mathbf{D}_0+\mathbf{K}_z|}
{|\mathbf{D}_0||\mathbf{D}_1+\mathbf{K}_z||\vd_2+\vk_z|}
\end{array}
\right\}.
\end{equation}
\end{theorem}
We show that if the distortion constraints
$(\vd_1,\;\vd_2,\;\vd_0)$ satisfy $ \vd_0 + \vk_x - \vd_1 - \vd_2
\succ \mathbf{0}$ and $\vd_0^{-1} + \vk_x^{-1} -
\vd_1^{-1}-\vd_2^{-1} \succ \mathbf{0}$, we can get the optimizing
$\vk_z$ by solving a matrix Riccati equation. An illustration of
the rate region is shown in Figure~\ref{fig:region}. In this case,
if we let $\vk_{w_l} = [\vd_l^{-1}-\vk_x^{-1}]^{-1}$ for
$l=0,1,2$, then the optimizing $\vk_z$ is
\[
\vk_z = \vk_x(\vk_x-\va^*)^{-1}\vk_x-\vk_x,
\]
where
\[
\va^* =
(\vk_{w_1}-\vk_{w_0})^{\frac{1}{2}}\left[(\vk_{w_1}-\vk_{w_0})^{-\frac{1}{2}}
(\vk_{w_2}-\vk_{w_0})(\vk_{w_1}-\vk_{w_0})^{-\frac{1}{2}}\right]^{\frac{1}{2}}(\vk_{w_1}-\vk_{w_0})^{\frac{1}{2}}
-\vk_{w_0}.
\]
Letting $R_{sum}$  denote the optimal sum rate,  the two corner
points in Figure~\ref{fig:region} are
$$
 B_1 =
\left(\frac{1}{2}\log\frac{|\vk_x|}{|\vd_1|},
R_{sum}-\frac{1}{2}\log\frac{|\vk_x|}{|\vd_1|}\right), \quad \mbox{and}$$
$$
B_2 = \left(R_{sum}-\frac{1}{2}\log\frac{|\vk_x|}{|\vd_2|},\frac{1}{2}\log\frac{|\vk_x|}{|\vd_2|}\right).$$


\section{Lower Bound}\label{sec:outerbound}
\label{sec3}
By fairly procedural steps,  we have the following lower bound to the
 sum rate of the multiple descriptions:
\begin{equation}
\begin{split}
n\sum\limits_{l=1}^LR_l  \ge & \sum\limits_{l=1}^LH(C_l) =
\sum\limits_{l=1}^LH(C_l)-H(C_1,\;\dots,\;C_L|\vx^n)  \\
 = & \sum\limits_{l=1}^LH(C_l)-H(C_1,\cdots,C_L) +
H(C_1,\;\dots,\;C_L) - H(C_1,\;\dots,\;C_L|\vx^n)  \\
 = & I(C_1;C_2;\dots;C_L) +
I(C_1,\;\dots,\;C_L; \vx^n),
\end{split}
\end{equation}
where we have defined
\[
I(C_1;C_2;\dots;C_L) \df
\sum\limits_{l=1}^LH(C_l)-H(C_1,\;\dots,\;C_L) =
\sum\limits_{l=2}^LI(C_l;C_1\dots C_{l-1}),
\]
and called it the symmetric mutual information between
$C_1,\;\dots,\;C_L$. Note that $I(C_1;C_2;\dots;C_L) \geq 0$ and is also well
defined even when $C_1,\;\dots,\;C_L$ are continuous random variables. Our
main result is  the
following information theoretic inequality which gives a lower
bound to the sum of symmetric mutual information between
$(C_1,C_2,\dots,C_L)$ and mutual information between
$C_1,C_2,\dots,C_L$ and $\vx^n$ for given covariance constraints.
\begin{lemma}\label{lemma:inform}
Let $\vx^n = (\vx[1],\;\dots,\;\vx[n])$, where $\vx[m]$'s are
i.i.d.\  $\mathcal{N}(\mathbf{0},\vk_x)$ Gaussian random vectors for
$m=1,\; \dots,\; n$. Let $C_1,\;\dots,\;C_L$ be random variables
jointly distributed with $\vx^n$. Let
$\hat{\vx}_0^n = \meane[\vx^n|C_1,\;\dots,\;C_L]$ and
$\hat{\vx}_l^n = \meane[\vx^n|C_l]$ for $l=1,\;\dots,\;L$. Given
positive definite matrices $\vd_1,\;\dots,\;\vd_L,\;\vd_0$, if
\begin{equation}\label{eq:lemmacov}
\begin{split}
\frac{1}{n}\sum\limits_{m=1}^n \meane[(\vx[m] -
\hat{\vx}_l[m])^t(\vx[m] - \hat{\vx}_l[m])] & \preccurlyeq
\mathbf{D}_l,
\quad  l=1,\;\dots,\;L, \\
\frac{1}{n}\sum\limits_{m=1}^n \meane[(\vx[m] -
\hat{\vx}_0[m])^t(\vx[m] - \hat{\vx}_0[m])] & \preccurlyeq
\mathbf{D}_0,
\end{split}
\end{equation}
then
\begin{equation}\label{eq:inequality}
I(C_1;C_2;\dots;C_L)+ I(C_1,\;\dots,\;C_L;
\vx^n) \ge \sup_{\vk_z \succ \mathbf{0}}
\frac{n}{2}\log\frac{|\vk_x||\vk_x+\mathbf{K}_z|^{(L-1)}|\mathbf{D}_0+\mathbf{K}_z|}
{|\mathbf{D}_0|\prod\limits_{l=1}^L|\mathbf{D}_l+\mathbf{K}_z|}.
\end{equation}
Furthermore, there exists a jointly Gaussian distribution of
$(C_1,\dots,C_L,\vx^n)$ such that the inequality in
\eqref{eq:inequality} is tight.
\end{lemma}

This is a fundamental information-theoretic inequality which
involves only the joint distribution\footnote{This inequality holds
even when $C_1,C_2,\dots,C_L$ are  not simply functions of $\vx^n$
and can also  be continuous random variables.} between
$C_1,C_2,\dots,C_L$ and $\vx^n$ and bounds on mean square error
estimation of $\vx^n$ from $C_1,C_2,\dots,C_L$; we delegate the
proof of this result to Appendix~\ref{app:inform}. We can now use
Lemma \ref{lemma:inform} to derive a lower bound to the sum rate
\begin{equation}\label{eq:outer}
\sum\limits_{l=1}^LR_l  \ge \sup_{\vk_z \succ \mathbf{0}}
\frac{1}{2}\log\frac{|\vk_x||\vk_x+\mathbf{K}_z|^{(L-1)}|\mathbf{D}_0+\mathbf{K}_z|}
{|\mathbf{D}_0|\prod\limits_{l=1}^L|\mathbf{D}_l+\mathbf{K}_z|}.
\end{equation}

By letting $L=1$ in the lemma above, we can derive a simple lower
bound to the  rate of the individual descriptions as well:
\begin{equation}\label{eq:individualbound}
\begin{split}
R_l & \ge \frac{1}{n}H(C_l) = \frac{1}{n}\big(H(C_l)-H(C_l|\vx^n)\big) \\
    & = \frac{1}{n} I(\vx^n;C_l)  \\
    & \ge \frac{1}{2}\log\frac{|\vk_x|}{|\mathbf{D}_l|}, \quad
    \listl.
\end{split}
\end{equation}
This bound is actually the point-to-point rate-distortion function
for individual receivers, since each individual receiver only
faces a point-to-point compression problem.

Note that for any positive definite $\vk_z$,
\[
\frac{1}{2}\log\frac{|\vk_x||\vk_x+\mathbf{K}_z|^{(L-1)}|\mathbf{D}_0+\mathbf{K}_z|}
{|\mathbf{D}_0|\prod\limits_{l=1}^L|\mathbf{D}_l+\mathbf{K}_z|}
\]
is a lower bound to the sum rate of the multiple descriptions. Two
special choices of $\vk_z$ are of particular interest:
\begin{itemize}
\item Letting $\vk_z = \epsilon \vi$ and 0 $\epsilon \rightarrow
0^+$, we have the following lower bound:
\begin{equation}\label{eq:kz0}
\displaystyle \sum\limits_{l=1}^LR_l \ge
\frac{1}{2}\log\frac{|\vk_x|^L}{|\vd_1|\dots|\vd_L|}.
\end{equation}
This bound is actually the summation of the bounds on the individual
rates.

\item Letting some eigenvalues of $\mathbf{K}_z$ goes to infinity, we
have the following lower bound:
\begin{equation}\label{eq:kzinf}
\displaystyle \sum\limits_{l=1}^LR_l \ge
\frac{1}{2}\log\frac{|\vk_x|} {|\mathbf{D}_0|}.
\end{equation}
This bound is the point-to-point rate-distortion function when we
only have the central distortion constraint.
\end{itemize}
 We will see later that for some distortion constraints
$(\vd_1,\;\dots,\;\vd_L,\;\vd_0)$, \eqref{eq:kz0} and
\eqref{eq:kzinf} can be tight.
\section{Upper Bound}\label{sec:innerbound}
\label{sec4}

In the previous section we gave a lower bound to the sum rate. Now
we give a upper bound to the sum rate by using the jointly
Gaussian description scheme described in Section
\ref{sec:sumrate}.

\subsection{Jointly Gaussian Multiple Description Scheme}

First we give a sketch of the achievable rate region by using
jointly Gaussian description scheme. Given the source sequence
$\vx^n$, as long as we can find a combination of codewords
$(\vu^n_1, \; \dots, \; \vu^n_L)$ that are jointly typical with
$\vx^n$, all the receivers can generate reproduction sequences that
satisfy their given distortion constraints. An intuitive way to
understand \eqref{eq:innerboundinu} is the following: since
$(\vu^n_1, \; \dots, \; \vu^n_L)$ are jointly typical with
$\vx^n$, then for any $S \subseteq \{1, \; \dots, \; L\}$, we have
that $\vu^n_l, l \in S$ are jointly typical with $\vx^n$. Now the
probability that a randomly generated combination of codewords
 $\vu^n_l, l \in S$ are jointly typical with $\vx^n$ is roughly
 \[
 \frac{e^{nh(\vu_l, l \in S|
 \vx)}}{\prod\limits_{l \in S}e^{nh(\vu_l)}},
 \]
and the number of possible combination of codewords $\vu^n_l, l
\in S$ are $\prod\limits_{l \in S}e^{nR_l}$. Thus, as long as
\begin{equation}\label{eq:raterandom}
\sum\limits_{l \in S}R_l \ge \sum\limits_{l \in S}h(\vu_l) -
h(\vu_l, l \in S|\vx),
\end{equation}
we can find a combination of codewords $\vu^n_l, l \in S$ that are
jointly typical with $\vx^n$. Rigorously speaking, we need to show that as
long as \eqref{eq:raterandom} is satisfied, then for any given
source sequence $\vx^n$ we can find a combination of codewords
$(\vu^n_1, \; \dots, \; \vu^n_L)$ such that $\vu^n_l, l \in S$ are
jointly typical with $\vx^n$ for all $S \subseteq \{1, \; \dots,
\; L\}$. The {\it second moment method}\cite{Alon00} is commonly
used to address this aspect, and a proof can be found in
\cite{Venkat03}.

Evaluating \eqref{eq:raterandom} based on the jointly Gaussian
distribution of $\vx$ and $\vu_1,\;\dots,\;\vu_L$, we get that all
the rate tuples $(R_1,\;\dots,\;R_L)$ satisfying
\begin{equation}
\sum\limits_{l \in S}R_l \ge \sum\limits_{l \in S}h(\vu_l) -
h(\vu_l, l \in S|\vx) = \frac{1}{2}\log\frac{\prod\limits_{l \in
S}|\vk_x+\vk_{w_l}|}{|\vk_{w_S}|}, \quad \forall S \subseteq \{1,
\; \dots, \; L\}
\end{equation}
are achievable by the jointly Gaussian description scheme. In
particular, we have that the achievable sum rate is
\begin{equation}\label{eq:achievablerate}
\sum\limits_{l=1}^Lh(\vu_l)-h(\vu_1,\;\dots,\;\vu_L|\vx) =
\frac{1}{2}\log\frac{\prod\limits_{l=1}^L|\vk_x+\vk_{w_l}|}{|\vk_w|}.
\end{equation}
The resulting distortions $(\vd^*_1,\;\dots,\;\vd^*_L,\;\vd^*_0)$
by using jointly Gaussian description scheme can be calculated as
\begin{equation}\label{eq:distortion2}
\begin{split}
\vd^*_l = & \cov[\vx|\vu_l] = [\vk_x^{-1} + \mathbf{K}_{w_l}^{-1}]^{-1}, \quad l=1,\;\dots,\; L ,  \\
\vd^*_0 = & \cov[\vx|\vu_1,\;\dots,\;\vu_L] = [\vk_x^{-1} +
(\vi,\; \dots, \; \vi)\mathbf{K}_w^{-1}(\vi,\; \dots, \;
\vi)^t]^{-1}.
\end{split}
\end{equation}

\subsection{Combinatorial Property of the Achievable Region}

The achievable region given in \eqref{eq:raterandom} has useful
combinatorial properties; in particular it belongs to the class of  {\it
contra-polymatroids}\cite{Welsh}. Certain rate regions of the multiple
access channel \cite{Tse98} and distributed source coding problems
\cite{Viswanath04} are also known to have this specific combinatorial
property. To see this, let
\[
\phi(S) \overset{\text{def}}{=} \sum\limits_{l \in S}h(\vu_l) -
h(\vu_l, l \in S|\vx), \quad S \subseteq \{1, \; \dots, \; L\}.
\]
We can readily verify that
\begin{equation}
\begin{split}
\phi(S\cup\{t\}) & \ge \phi(S), \quad \forall t \in \{1, \; \dots,
\; L\}, \\
\phi(S\cup T)+\phi(S\cap T) & \ge \phi(S) + \phi(T).
\end{split}
\end{equation}
By definition, we conclude that  the achievable rate region of a
jointly Gaussian multiple description scheme is a contra-polymatroid.
The key advantage of this combinatorial propety is that we can exactly
characterize the vertices of the achievable rate region
\eqref{eq:raterandom}. Letting $\pi$ to be a
permutation on $\{1, \; \dots, \; L\}$, define
\[
b_i^{(\pi)} \overset{\text{def}}{=}
\phi(\{\pi_1,\pi_2,\dots,\pi_i\})-\phi(\{\pi_1,\pi_2,\dots,\pi_{i-1}\}),
\quad i=1, \; \dots, \; L,
\]
and $\mathbf{b}^{(\pi)} = \left(b_1^{(\pi)}, \dots,
b_L^{(\pi)}\right)$. Then the $L!$ points $\{\mathbf{b}^{(\pi)},
\pi \text{ a permutation}\}$ are the vertices of the
contra-polymatroid \eqref{eq:raterandom}.


\subsection{Comparison of Upper Bound and the Lower Bound}

Our goal is to  show that the jointly Gaussian description scheme
achieves the lower bound to the  sum rate. In general it does not
seem facile  to do a direct calculation and comparison. We forgo
this strategy and, instead, provide an alternative
characterization of the achievable sum rate which is much easier
to compare with the lower
bound. 

Similar to the derivation of the lower bound (in Appendix~\ref{app:inform}),
we consider an
$\mathcal{N}(0, \mathbf{K}_z)$ Gaussian random vector $\vz$,
independent of $\vx$ and all $\vw_l$'s. Defining $\vy=\vx+\vz$,
we have the following achievable sum rate:
\begin{eqnarray}\label{eq:suminnerbound25}
\sum\limits_{l=1}^LR_l & = &
\sum\limits_{l=1}^Lh(\vu_l)-h(\vu_1,\;\dots,\;\vu_L|\vx) \nonumber \\
& = & \sum\limits_{l=1}^Lh(\vu_l)-h(\vu_1,\;\dots,\;\vu_L) +
h(\vu_1,\;\dots,\;\vu_L) - h(\vu_1,\;\dots,\;\vu_L|\vx) \nonumber \\
& = & \sum\limits_{l=1}^Lh(\vu_l)-h(\vu_1,\cdots,\vu_L) +
I(\vu_1,\;\dots,\;\vu_L; \vx) \nonumber \\
& \overset{(a)}{\ge} &
\sum\limits_{l=1}^Lh(\vu_l)-h(\vu_1,\cdots,\vu_L) +
I(\vu_1,\;\dots,\;\vu_L; \vx) -
\left(\sum\limits_{l=1}^Lh(\vu_l|\vy)-h(\vu_1,\;\dots,\;\vu_L|\vy)\right) \nonumber \\
& = & \sum\limits_{l=1}^L\big(h(\vy)-h(\vy|\vu_l)\big)-h(\vy)
+h(\vy|\vu_1,\;\dots,\;\vu_L) + h(\vx)-h(\vx|\vu_1,\;\dots,\;\vu_L) \nonumber \\
& = & h(\vx)+(L-1)h(\vy)-\sum\limits_{l=1}^Lh(\vy|\vu_l)+
h(\vy|\vu_1,\;\dots,\;\vu_L)-h(\vx|\vu_1,\;\dots,\;\vu_L)
\nonumber \\
& = &
\frac{1}{2}\log\frac{\Big|\vk_x\Big|\Big|\vk_x+\mathbf{K}_z\Big|^{(L-1)}\Big|\cov[\vx|\vu_1,\;\dots,\;\vu_L]+\mathbf{K}_z\Big|}
{\Big|\cov[\vx|\vu_1,\;\dots,\;\vu_L]\Big|\prod\limits_{l=1}^L\Big|\cov[\vx|\vu_l]+\mathbf{K}_z\Big|},
\end{eqnarray}
where the last step is from a procedural Gaussian MMSE calculation.

Note that if we have
\begin{equation}\label{eq:independent}
\sum\limits_{l=1}^Lh(\vu_l|\vy)-h(\vu_1,\;\dots,\;\vu_L|\vy) = 0,
\end{equation}
then (a) in \eqref{eq:suminnerbound25} is actually an equality.
Thus, if our choice of $\mathbf{K}_w$ and $\mathbf{K}_z$ satisfy
the following two conditions:
\begin{itemize}
\item  \eqref{eq:independent} is true.

\item distortion constraints are met with equality, i.e.,
\begin{equation}\label{eq:distortion}
\begin{split}
& \cov[\vx|\vu_l]  = \mathbf{D}_l, \quad l=1,\;\dots,\; L, \\
& \cov[\vx|\vu_1,\;\dots,\;\vu_L]  = \mathbf{D}_0,
\end{split}
\end{equation}
\end{itemize}
then the upper bound matches the lower bound and we have characterized the
 sum rate. In the following we examine under what
circumstances the above two conditions are true.

First, we give a necessary and sufficient condition for
\eqref{eq:independent} to be true, delegating the proof to
Appendix~\ref{app:nec_suff_condition}.
\begin{proposition}
\label{prop:nec_suff_condition}
There exists some choice of positive definite $\vk_z$ such that
\eqref{eq:independent} is true if and only if $\vk_w$, the
covariance matrix of $(\vw_1,\; \cdots, \; \vw_L)$, takes the
following form
\begin{equation}\label{eq:kw}
\mathbf{K}_w =
        \begin{pmatrix}
                \mathbf{K}_{w_1} & -\mathbf{A} & -\mathbf{A} & \dots & -\mathbf{A} \\
                -\va & \mathbf{K}_{w_2} & -\va & \dots & -\va \\
                \hdotsfor{5} \\
                -\va & \dots & -\va & \mathbf{K}_{w_{L-1}} & -\va \\
                -\va & \dots & -\va & -\va & \mathbf{K}_{w_L}
       \end{pmatrix},
\end{equation}
where $ \mathbf{0} \prec \va \prec \vk_x$.
\end{proposition}

Next, we look at the conditions for \eqref{eq:distortion} to be
true. From \eqref{eq:distortion2}, we have
\begin{equation}\label{eq:distortion1}
\begin{split}
\mathbf{D}_l^{-1} & = \cov[\vx|\vu_l]^{-1}  = \vk_x^{-1} + \mathbf{K}_{w_l}^{-1}, \quad l=1,\;\dots,\; L \\
\mathbf{D}_0^{-1} & = \cov[\vx|\vu_1,\;\dots,\;\vu_L]^{-1}  =
\vk_x^{-1} + (\vi,\; \dots, \; \vi)\mathbf{K}_w^{-1}(\vi,\; \dots,
\; \vi)^t.
\end{split}
\end{equation}
$(\vi,\; \vi, \; \dots, \;
\vi)\mathbf{K}_w^{-1}(\vi,\; \vi, \; \dots, \; \vi)^t$, is calculated in the  following lemma; the proof is available in Appendix~\ref{app:inverse}.
\begin{lemma}\label{lemma:inverse}
Let
\[
\vk_w = \begin{pmatrix}
                \vk_{w_1} & -\va & -\va & \dots & -\va \\
                -\va & \vk_{w_2} & -\va & \dots & -\va \\
                \hdotsfor{5} \\
                -\va & \dots & -\va & \vk_{w_{L-1}} & -\va \\
                -\va & \dots & -\va & -\va & \vk_{w_L}.
       \end{pmatrix}.
\]
If $\vk_w \succ \mathbf{0}$ and $\va \succeq \mathbf{0}$, then
\[
(\vi,\; \vi, \; \dots, \; \vi)\mathbf{K}_w^{-1}(\vi,\; \vi, \;
\dots, \; \vi)^t = \left[\left(\sum\limits_{l=1}^L
(\vk_{w_l}+\va)^{-1}\right)^{-1}-\va\right]^{-1}.
\]
\end{lemma}
Using this lemma, from \eqref{eq:distortion1} we arrive at
\begin{equation}
\left[(\vd_0^{-1}-\vk_x^{-1})^{-1}+\va\right]^{-1} =
\sum\limits_{l=1}^L\left[(\vd_l^{-1}-\vk_x^{-1})^{-1}+\va\right]^{-1}.
\end{equation}
Defining
\begin{equation}
\vk_{w_0} = (\vd_0^{-1}-\vk_x^{-1})^{-1},
\end{equation}
\eqref{eq:distortion1} is equivalent to
\begin{equation}\label{eq:core}
\left[\vk_{w_0}+\va\right]^{-1}  =
\sum\limits_{l=1}^L\left[\vk_{w_l}+\va\right]^{-1}.
\end{equation}
Thus, if there exists a positive definite solution $\va$ to
\eqref{eq:core}, and the corresponding $\vk_w$ is positive definite,
then the distortion constraints are met with equality, i.e.,
\eqref{eq:distortion} holds. It turns out that as long as $\va$ is a solution to
\eqref{eq:core}, the resulting $\vk_w$ is always positive definite; we state this formally below, delegating the proof to  Appendix \ref{app:positivedefinite}.
\begin{lemma}\label{lemma:positivedefinite}
If for some $\vk_{w_0} \succ \mathbf{0}$ and $\va \succ
\mathbf{0}$ \eqref{eq:core} is true, then the covariance matrix
$\vk_w$ defined in \eqref{eq:kw} is positive definite.
\end{lemma}
We summarize the state of affairs  in the following theorem.
\begin{theorem}\label{th:main}
Given distortion constraints $(\vd_1, \; \dots, \; \vd_L, \vd_0)$,
let
\begin{equation}
\vk_{w_l} = (\vd_l^{-1}-\vk_x^{-1})^{-1}, \quad l=0, \; 1, \;
\dots, \; L.
\end{equation}
If there exists an solution $\va^*$ to \eqref{eq:core} and
$\mathbf{0} \prec \va^* \prec \vk_x$, then the jointly Gaussian
description scheme with $\vk_w$ defined in \eqref{eq:kw} with $\va
= \va^*$ achieves the optimal sum rate, and the optimal $\vk_z$
for lower bound \eqref{eq:outer} is $\mathbf{K}_z =
\vk_x(\vk_x-\va^*)^{-1}\vk_x-\vk_x$.
\end{theorem}

Thus we show that if the given distortion constraints $(\vd_1, \;
\dots, \; \vd_L, \vd_0)$ satisfy the condition for Theorem
\ref{th:main}, then the jointly Gaussian description scheme
achieves the optimal sum rate and we can calculate the optimal
$\vk_w$ by solving a matrix equation. However, for arbitrarily
given distortion constraints, \eqref{eq:core} may not have a
solution $\va^*$ such that $\mathbf{0} \prec \va^* \prec \vk_x$.
In this case, we can show that there exists a jointly Gaussian
description scheme that achieves the sum rate lower bound, and
resulting in distortions $(\vd^*_1, \; \dots, \; \vd^*_L,
\vd^*_0)$ such that $\vd^*_l \preccurlyeq \vd_l$ for
$l=0,1,\dots,L$. In the following we first study the relatively
simpler case of scalar Gaussian source, and then move to  discuss
the vector Gaussian source.

\section{Scalar Gaussian Source}\label{section:scalar}
\label{sec5}
Here we suppose that  the  information source is an i.i.d.\ sequence
 of $\mathcal{N}(0,\sigma_x^2)$ scalar Gaussian random
variables. Let individual distortion constraints be
$(d_1,\;\dots,\;d_L)$ and the central distortion constraints be
$d_0$, where $0 < d_0 < d_l < \sigma_x^2$ for $l=1,\; \dots, \;
L$. We consider the jointly Gaussian description scheme with the
following covariance matrix for $w_1, \; \dots, \; w_l$.
\begin{equation}\label{eq:scalarkw}
\vk_w = \begin{pmatrix}
                \sigma_1^2 & -a & -a & \dots & -a \\
                -a & \sigma_2^2 & -a & \dots & -a \\
                \hdotsfor{5} \\
                -a & \dots & -a & \sigma_{L-1}^2 & -a \\
                -a & \dots & -a & -a & \sigma_L^2
       \end{pmatrix}.
\end{equation}

Consider the condition for Theorem \ref{th:main} to hold: to meet
the individual distortion constraint with equality, we need
\begin{equation}\label{eq:sigmal}
\sigma_l^2 = (d_l^{-1}-\sigma_x^{-2})^{-1} =
\frac{d_l\sigma_x^2}{\sigma_x^2-d_l}, \quad l=1,\;\dots,\;L.
\end{equation}
Let
\begin{equation}
\sigma_0^2 \overset{\text{def}}{=} (d_0^{-1}-\sigma_x^{-2})^{-1} =
\frac{d_0\sigma_x^2}{\sigma_x^2-d_0},
\end{equation}
we need
\begin{equation}
\left[\sigma_0^2+a\right]^{-1}  =
\sum\limits_{l=1}^L\left[\sigma_l^2+a\right]^{-1}
\end{equation}
to have a solution $a^* \in (0, \sigma_x^2)$, to meet the central
distortion constraint with equality.  Towards this, define
\begin{equation}
f(a)\overset{\text{def}}{=} \frac{1}{\sigma_0^2+a} -
\sum\limits_{l=1}^L\frac{1}{\sigma_l^2+a},
\end{equation}
and we have
\begin{equation}
\begin{split}
f(0) & = \frac{1}{\sigma_0^2}-
\sum\limits_{l=1}^L\frac{1}{\sigma_l^2} =
\frac{1}{d_0}+\frac{L-1}{\sigma_x^2}-\sum\limits_{l=1}^L\frac{1}{d_l}, \\
f(\sigma_x^2) & = \frac{1}{\sigma_0^2+\sigma_x^2}-
\sum\limits_{l=1}^L\frac{1}{\sigma_l^2+\sigma_x^2} =
\frac{1}{\sigma_x^4}\left(\sum\limits_{l=1}^Ld_l-d_0-(L-1)\sigma_x^2\right).
\end{split}
\end{equation}
Using induction, we can show that
\begin{equation}
\left(\sum\limits_{l=1}^L\frac{1}{d_l}-\frac{L-1}{\sigma_x^2}\right)^{-1}
\ge \sum\limits_{l=1}^Ld_l-(L-1)\sigma_x^2.
\end{equation}
Thus we have
\begin{equation*}
\begin{split}
f(0) \le 0 & \Rightarrow f(\sigma_x^2) \le 0, \\
f(\sigma_x^2) \ge 0 & \Rightarrow f(0) \ge 0.
\end{split}
\end{equation*}
Then given distortions $(d_1, \; \dots, \; d_L, \; d_0)$, $f(0)$
and $f(\sigma_x^2)$ falls into the following three cases.


{\bf Case 1:} $f(0) > 0$ and $f(\sigma_x^2) <0$.

In this case, since $f(a)$ is a continuous function, there exists
an $a^* \in (0,\sigma_x^2)$ such that $f(a^*) = 0$. In this case
the condition for Theorem \ref{th:main} holds and from Theorem
\ref{th:main} we know that jointly Gaussian description scheme
with covariance matrix for $w_1, \; \dots, \; w_l$ being
\eqref{eq:scalarkw} with $a = a^*$ achieves the optimal sum rate.

{\bf Case 2:} $f(0) \le 0$. Alternatively,
$\frac{1}{d_0}+\frac{L-1}{\sigma_x^2}-\sum\limits_{l=1}^L\frac{1}{d_l}
\le 0$.

In this case, the condition for Theorem \ref{th:main} does not
hold. But the jointly Gaussian description scheme can still achieve
the sum rate. To see this,  choosing $a=0$ in $\mathbf{K}_w$ we
can meet individual distortions with equality and get a central
distortion $d_0'$. From \eqref{eq:distortion1} we have
\begin{equation}
\begin{split}
\frac{1}{d_0'} & = \frac{1}{\sigma_x^2}+ (1 \; 1 \;  \dots \;
1)K_w^{-1} (1 \; 1 \; \dots
\; 1)^t \\
& = \frac{1}{\sigma_x^2}+\sum\limits_{l=1}^L\frac{1}{\sigma_l^2} = \sum\limits_{l=1}^L\frac{1}{d_l} - \frac{L-1}{\sigma_x^2} \\
& \geq \frac{1}{d_0}.
\end{split}
\end{equation}
Hence we have achieved distortion $(d_1,\dots,d_L,d'_0)$ where $d_0'
\le d_0$, and from \eqref{eq:achievablerate} the achievable sum
rate is
\begin{equation}
\sum\limits_{l=1}^LR_l  \ge
\frac{1}{2}\log\frac{\sigma_x^{2L}}{d_1d_2\cdots d_{L}},
\end{equation}
which equals  the sum of our bounds on individual rates.

{\bf Case 3:} $f(\sigma_x^2) \ge 0$, Alternatively,  $
\sum\limits_{l=1}^Ld_l - d_0 - (L-1)\sigma_x^2 \geq 0 $.

In this case, the conditions for Theorem \ref{th:main} do not
hold as well. But the jointly Gaussian description strategy still
achieves the sum rate. To see this,
note that we can find a $d'_L$ such that $0 < d'_L \le d_L$ and
\begin{equation}\label{eq:s3}
\sum\limits_{l=1}^{L-1}d_l + d'_L- d_0 - (L-1)\sigma_x^2 =  0,
\end{equation}
and we choose $a = \sigma_x^2$, $\sigma_l^2 =
(d_l^{-1}-\sigma_x^{-2})^{-1}$ for $l=1,\cdots,L-1$, and
$\sigma_L^2 = ({d'}_L^{-1}-\sigma_x^{-2})^{-1}$ in $K_w$. Defining
$\sigma_0^2 = (d_0^{-1}-\sigma_x^{-2})^{-1}$,  \eqref{eq:s3}
is equivalent to the following equation:
\begin{equation}\label{eq:s4}
\left[\sigma_0^2+\sigma_x^2\right]^{-1}  =
\sum\limits_{l=1}^L\left[\sigma_l^2+\sigma_x^2\right]^{-1}.
\end{equation}
From Lemma \ref{lemma:positivedefinite}, our choice of $K_w$ is
positive definite. Thus  the resulting distortions are $(d_1, \;
\dots, \; d_{L-1}, \; d'_L, d_0)$, where $0 < d'_L \le d_L$.

Using the determinant equation \begin{equation}
\begin{vmatrix}
                \sigma_1^2 & -\sigma_x^2 & -\sigma_x^2 &  -\sigma_x^2 & \dots & -\sigma_x^2\\
                -\sigma_x^2 & \sigma_2^2 & -\sigma_x^2 & -\sigma_x^2 & \dots & -\sigma_x^2 \\
                -\sigma_x^2 & -\sigma_x^2 & \sigma_3^2 & -\sigma_x^2 & \dots & -\sigma_x^2 \\
                \hdotsfor{6} \\
                -\sigma_x^2 & \dots & -\sigma_x^2 & -\sigma_x^2 & \sigma_{L-1}^2 & -\sigma_x^2 \\
                -\sigma_x^2 & \dots & -\sigma_x^2 & -\sigma_x^2 & -\sigma_x^2 & \sigma_L^2
       \end{vmatrix}
=
\Big(1-\sum\limits_{l=1}^L\frac{\sigma_x^2}{\sigma_l^2+\sigma_x^2}\Big)\prod\limits_{l=1}^L(\sigma_l^2+\sigma_x^2)
\end{equation}
and \eqref{eq:s4}, we have an   achievable  sum rate
\begin{equation}
\sum\limits_{l=1}^LR_l  =  \frac{1}{2}\log
\frac{\sigma_x^2}{d_0}.
\end{equation}
We conclude that  in this case the point-to-point rate-distortion
bound  for the central receiver is achievable.

In summary,  we have shown that the jointly Gaussian description
scheme achieves the lower bound on the sum rate. Further,  the sum
rate can be calculated either trivially (by choosing $a^*=0$ in
case II or $a^*=1$ in case III) or by solving a polynomial
equation in a single variable (case I).

\section{Vector Gaussian Source}
\label{sec6} The essence of our proof of the optimality of jointly
Gaussian description scheme for scalar Gaussian sources is the use
of the {\em
  intermediate value theorem} for scalar continuous
functions. However, there is no natural extension of this theorem
for vector valued functions. To avoid this problem, we first
explicitly solve the two description problem and characterize the
optimality of jointly Gaussian description scheme. Next, we show
that the jointly Gaussian description scheme is optimal for $L
 \ge 2$ by showing an equivalence of certain optimization
 problems. In the last part of this section, we show that
the jointly Gaussian description strategy can achieve the optimal
 rate region for the two description problem.

\subsection{Explicit Solutions for Some Cases of Two Description
Problem}\label{sec:vector2}

With only  two descriptions, we can explicitly solve
\eqref{eq:core}, thus  generalizing the corresponding solution for
the scalar  Gaussian source, derived in \cite{Ozarow80}.

Suppose the distortion constraints are denoted by  $(\vd_1,\;
\vd_2, \; \vd_0)$ and let
\[
\vk_w = \begin{pmatrix}
                \vk_{w_1} & -\va^* \\
                -\va^* & \vk_{w_2}
       \end{pmatrix}.
\]
We now solve \eqref{eq:distortion1}, which is equivalent to
\eqref{eq:core}, for $\vk_{w_1}$, $\vk_{w_2}$ and $\va^*$. From
\eqref{eq:distortion1} we get
\begin{equation}
\vk_{w_l} = (\vd_l^{-1}-\vk_x^{-1})^{-1}, \quad l=1,\;2,
\end{equation}
and
\begin{equation}
\vd_0^{-1} = \vk_x^{-1}+ (\vi \; \vi)\vk_w^{-1}(\vi \; \vi)^t.
\end{equation}
Expanding out $\vk_w^{-1}$ using Lemma \ref{lemma:blockinverse} in
Appendix \ref{app:matrix}, we get
\begin{equation}
\vd_0^{-1}-\vk_x^{-1} =
\vk_{w_1}^{-1}+(\vi+\vk_{w_1}^{-1}\va^*)(\vk_{w_2}-\va^*\vk_{w_1}^{-1}\va^*)^{-1}(\vi+\va^*\vk_{w_1}^{-1}).
\end{equation}
Taking inverse on both sides, we have
\begin{equation}\label{eq:47}
(\vd_0^{-1} - \vk_x^{-1})^{-1} =
\vk_{w_1}-(\vk_{w_1}+\va^*)(\vk_{w_1}+\vk_{w_2}+2\va^*)^{-1}(\vk_{w_1}+\va^*).
\end{equation}
Defining $\vk_{w_0}$ as
\begin{equation}
\vk_{w_0} \overset{\text{def}}{=} [\vd_0^{-1} - \vk_x^{-1}]^{-1} ,
\end{equation}
\eqref{eq:47} is equivalent to
\begin{equation}\label{eq:49}
\vk_{w_1}-\vk_{w_0} =
(\vk_{w_1}+\va^*)(\vk_{w_1}+\vk_{w_2}+2\va^*)^{-1}(\vk_{w_1}+\va^*).
\end{equation}
Defining
\[
\mathbf{X} \overset{\text{def}}{=} \vk_{w_1}+\va^*,
\]
\eqref{eq:49} is equivalent to
\begin{equation}
\vk_{w_1}-\vk_{w_0} =
\mathbf{X}(2\mathbf{X}+\vk_{w_2}-\vk_{w_1})^{-1}\mathbf{X},
\end{equation}
which is further equivalent to
\begin{equation}
\mathbf{X}(\vk_{w_1}-\vk_{w_0})^{-1}\mathbf{X} =
2\mathbf{X}+\vk_{w_2}-\vk_{w_1}.
\end{equation}
This is a version of the so-called {\em algebraic  Riccati equation};
the corresponding Hamiltonian is readily seen to be positive
semidefinite and we can even write down the following explicit solution:
\begin{equation}
\begin{split}
\mathbf{X}  = & \vk_{w_1}-\vk_{w_0}
\\
 & +(\vk_{w_1}-\vk_{w_0})^{\frac{1}{2}}\left[(\vk_{w_1}-\vk_{w_0})^{-\frac{1}{2}}
(\vk_{w_2}-\vk_{w_0})(\vk_{w_1}-\vk_{w_0})^{-\frac{1}{2}}\right]^{\frac{1}{2}}(\vk_{w_1}-\vk_{w_0})^{\frac{1}{2}}.
\end{split}
\end{equation}
Thus
\begin{equation}
\va^* =
(\vk_{w_1}-\vk_{w_0})^{\frac{1}{2}}\left[(\vk_{w_1}-\vk_{w_0})^{-\frac{1}{2}}
(\vk_{w_2}-\vk_{w_0})(\vk_{w_1}-\vk_{w_0})^{-\frac{1}{2}}\right]^{\frac{1}{2}}(\vk_{w_1}-\vk_{w_0})^{\frac{1}{2}}
-\vk_{w_0}.
\end{equation}
Now, if $\mathbf{0} \prec \va^* \prec \vk_x$ then we can appeal to
Theorem~\ref{th:main} and arrive at the explicit jointly Gaussian
description scheme parameterized by $\vk_w$ that achieves the sum
rate. Analogous to the scalar case (cf.\ \cite{Ozarow80}),  we
have the following sufficient condition for when this is true; the
proof is available in Appendix~\ref{app:twod}.
\begin{proposition}\label{prop:twod}
If the distortion constraints $(\vd_1,\;\vd_2,\;\vd_0)$ satisfy
\begin{equation}\label{eq:vdfor2}
\begin{split}
& \vd_0 + \vk_x - \vd_1 - \vd_2  \succ  \mathbf{0} \\
{\textup{and}} \quad &  \vd_0^{-1} + \vk_x^{-1} -
\vd_1^{-1}-\vd_2^{-1} \succ \mathbf{0},
\end{split}
\end{equation}
then $\mathbf{0} \prec \va^* \prec \vk_x$.
\end{proposition}
We now complete the proof by considering  the  cases that are not
covered by the conditions in Proposition~\ref{prop:twod}.
\begin{itemize}
\item When
\[
\vd_0^{-1} + \vk_x^{-1} - \vd_1^{-1}-\vd_2^{-1}  \preccurlyeq
\mathbf{0},
\]
we can choose $\va^* = \mathbf{0}$ to achieve the sum of
point-to-point individual rate-distortion functions. Thus in this
case, the sum rate is equal to this natural lower bound.
\item  When
\[
\vd_0 + \vk_x - \vd_1 - \vd_2 \preccurlyeq  \mathbf{0},
\]
we can choose $\va^* = \vk_x$ to achieve the point-to-point rate
distortion-function for central receiver, also a natural lower
bound.

\item When neither $\vd_0 + \vk_x - \vd_1 - \vd_2$ nor $\vd_0^{-1}
+ \vk_x^{-1} - \vd_1^{-1}-\vd_2^{-1}$ is positive or negative
semidefinite (this case cannot happen in the scalar case), we cannot
use Theorem~\ref{th:main}, and the trivial choice of $\va^* =
\mathbf{0}$ or $\va^* = \vk_x$ does not meet the lower bound. In
the next subsection we will address this case and prove that
the jointly Gaussian
description scheme indeed  achieves the lower bound on the sum rate
for $L \ge 2$.

\end{itemize}


If we let the source to be  scalar Gaussian, our result reduces to
Ozarow's solution of the two description problem for a scalar Gaussian
source\cite{Ozarow80}: this is because the last case described above
does not happen in the scalar case.

\subsection{Solutions for $L \ge 2$}\label{sec:vecl}
While we exactly characterized the optimal jointly Gaussian
description scheme and used this characterization in arguing that
it achieves the fundamental lower bound to the sum rate, such
exact calculations do not appear to be as immediate when $L>2$.
So, we eschew this somewhat brute-force approach and resort to a
more subtle proof  that involves exploring the structure of the
solution to an optimization problem. First, note that by a linear
transformation at the encoders and the decoders, we have the
following result on rate region for multiple description with
individual and central receivers.
\begin{proposition}
\begin{equation}
 R_*(\vk_x,\;\vd_1, \; \dots, \;, \vd_L, \;
\vd_0) = R_*(\vi,\;\vk_x^{-\frac{1}{2}}\vd_1\vk_x^{-\frac{1}{2}},
\; \dots, \;, \vk_x^{-\frac{1}{2}}\vd_L\vk_x^{-\frac{1}{2}}, \;
\vk_x^{-\frac{1}{2}}\vd_0\vk_x^{-\frac{1}{2}}).
\end{equation}
\end{proposition}
Thus, throughout this subsection we will suppose,  for notation
simplicity, that $\vk_x = \vi$.

Given distortion constraints $(\mathbf{D}_1,\; \dots \, \;
\mathbf{D}_L,\; \mathbf{D}_0)$, let
\begin{equation}
\vk_{w_l}  = (\vd_l^{-1}-\vi)^{-1}, \quad l=0, \; 1, \; \dots, \;
L,
\end{equation}
and define
\begin{eqnarray}
f(\va) \overset{\text{def}}{=} & \left[\vk_{w_0}+\va\right]^{-1} -
\sum\limits_{l=1}^L\left[\vk_{w_l}+\va\right]^{-1}, \\
F(\va) \overset{\text{def}}{=} & \log |\vk_{w_0}+\va| -
\sum\limits_{l=1}^L \log |\vk_{w_l}+\va|.
\end{eqnarray}
Note that
\begin{equation}
\frac{dF(\va)}{d\va} = f(\va).
\end{equation}
Consider the following optimization problem:
\begin{equation}\label{eq:optimal}
\max\limits_{\mathbf{0} \preccurlyeq \va \preccurlyeq
\vi}\quad\quad F(\va).
\end{equation}
Now, since $F(\va)$ is a continuous map and $\mathbf{0} \preccurlyeq
\va \preccurlyeq \vi$ is a compact set, there exists an optimal
solution $\va^*$ to \eqref{eq:optimal} where  $\va^*$ satisfies the
Karush-Kuhn-Tucker (KKT) conditions: there exist $\vl_1  \succcurlyeq
\mathbf{0}$ and $\vl_2  \succcurlyeq \mathbf{0}$ such that
\begin{eqnarray}
f(\va^*)+\vl_1-\vl_2 & = & \mathbf{0} \label{eq:kkt1}\\
\vl_1 \va^* &  = & \mathbf{0}\label{eq:kkt2} \\
\vl_2 (\va^*-\vi) & = & \mathbf{0}\label{eq:kkt3}.
\end{eqnarray}
Now  $\va^*$ falls into the following four cases.

{\bf Case 1:}  $\mathbf{0} \prec \mathbf{A}^* \prec \mathbf{I}$.
Alternatively,  0 and 1 are not eigenvalues of $\va^*$. In this
case, $\vl_1 = {\bf 0}$ and  $\vl_2 ={\bf 0}$; thus the KKT
conditions in \eqref{eq:kkt1} reduce to
\[
f(\va^*) = \mathbf{0}.
\]
Equivalently,
\begin{equation}
\left[\vk_{w_0}+\va^*\right]^{-1} =
\sum\limits_{l=1}^L\left[\vk_{w_l}+\va^*\right]^{-1}.
\end{equation}
From Theorem~\ref{th:main}, the jointly Gaussian description
scheme with covariance matrix for $\vw_1, \; \dots, \; \vw_L$
being
\begin{equation}\label{eq:kwvector}
\vk_w = \begin{pmatrix}
                \vk_{w_1} & -\va^* & -\va^* & \dots & -\va^* \\
                -\va^* & \vk_{w_2} & -\va^* & \dots & -\va^* \\
                \hdotsfor{5} \\
                -\va^* & \dots & -\va^* & \vk_{w_{L-1}} & -\va^* \\
                -\va^* & \dots & -\va^* & -\va^* & \vk_{w_L}
       \end{pmatrix}
\end{equation}
 achieves the lower bound to the  sum rate. Thus in this case, we have
 characterized the optimality of the jointly Gaussian
 description scheme parameterized by \eqref{eq:kwvector} in terms of
 achieving the sum rate.

{\bf Case 2:} $\mathbf{0} \preccurlyeq \va^* \prec \vi$. Alternatively,
 some eigenvalues of $\va^*$ are 0, but no eigenvalues of $\va^*$ are
 1. Thus $\vl_2 = {\bf 0}$ and the KKT conditions in \eqref{eq:kkt1}
 reduce to
\begin{equation}\label{eq:kkt1_again}
(\vk_{w_0}+\va^*)^{-1} -
\sum\limits_{l=1}^L(\vk_{w_l}+\va^*)^{-1}+\vl_1 = \mathbf{0},
\end{equation}
for some $\vl_1 \succcurlyeq \mathbf{0}$ satisfying $\vl_1 \va^*  =
\mathbf{0}$. The {\em key idea} now is to see that the distortion constraint
on the central receiver is too loose and we can in fact achieve a {\em
  lesser} distortion (in the sense of  positive semidefinite ordering)
for the same sum rate. We first identify this lower distortion:
defining
$$
\vk_{w_0}^*  = \left(\vk_{w_0}^{-1}+\vl_1\right)^{-1},
$$
consider the smaller distortion matrix on the central receiver
$$
\vd_0^*  = \left( {\vk_{w_0}^*}^{-1}+\vi\right)^{-1} = \left(\vi+
\vk_{w_0}^{-1}+\vl_1\right)^{-1} = (\vd_0^{-1} + \vl_1)^{-1} \prec
\vd_0.
$$
This new distortion matrix on the central receiver satisfies two
key properties, that we state as a lemma (whose proof is available
in Appendix \ref{app:enhanced0}).
\begin{lemma}\label{lemma:enhanced0}
\begin{eqnarray}
(\vk_{w_0}+\va^*)^{-1}+\vl_1 & = & (\vk_{w_0}^*+\va^*)^{-1},
\label{eq:enhanced_1}\\
\frac{|\vd_0+\vk_z|}{|\vd_0|} & = &
\frac{|\vd_0^*+\vk_z|}{|\vd_0^*|}.\label{eq:enhanced0_1}
\end{eqnarray}
\end{lemma}
Comparing \eqref{eq:kkt1_again} with \eqref{eq:enhanced_1}, we have
\begin{equation}\label{eq:vk}
\left[\vk^*_{w_0}+\va^*\right]^{-1} =
\sum\limits_{l=1}^{L}\left[\vk_{w_l}+\va^*\right]^{-1}.
\end{equation}
Now, the corresponding $\vk_z = (\vi-\va^*)^{-1}-\vi$ is singular.
If it hadnt been, then by Theorem~\ref{th:main} we could have
concluded that jointly  Gaussian description scheme achieves the
lower bound to the sum rate. We now address this  technical
difficulty.

 Our first observation is that  there exists $\delta >0$ such that for
 all $\epsilon \in
(0,\delta)$ we have $\mathbf{0} \prec \va+\epsilon \vi \prec \vi$,
and $0 \prec \vk_{w_0}^*-\epsilon \vi$, $\mathbf{0} \prec
\vk_{w_l}-\epsilon \vi$, and we can rewrite \eqref{eq:vk} as
\begin{equation}\label{eq:continues1}
\big[(\vk_{w_0}^*-\epsilon \vi)+(\va^*+\epsilon \vi)\big]^{-1} =
\sum\limits_{l=1}^L\big[(\vk_{w_l}-\epsilon \vi)+(\va^*+\epsilon
\vi)\big]^{-1}.
\end{equation}
Thus if the distortion constraints were
$(\vd_1(\epsilon), \; \dots , \; \vd_L(\epsilon), \;
\vd_0(\epsilon))$ with
\begin{equation*}
\begin{split}
\vd_l(\epsilon) &  = \left[(\vk_{w_l}-\epsilon
\vi)^{-1}+\vi\right]^{-1}, \quad \listl, \\
\vd_0(\epsilon) &  = \left[(\vk_{w_0}^*-\epsilon
\vi)^{-1}+\vi\right]^{-1},
\end{split}
\end{equation*}
then $ \va^*+\epsilon \vi$ is a solution to
\eqref{eq:continues1}. This situation corresponds to that discussed in
Case I; we can conclude that   sum rate for this modified distortion
multiple description problem is
\begin{equation}
\frac{1}{2}\log\frac{|\vi+\vk_z(\epsilon)|^{(L-1)}|\vd_0(\epsilon)+\vk_z(\epsilon)|}{|\vd_0(\epsilon)|\prod\limits_{l=1}^L
|\vd_l(\epsilon)+\vk_z(\epsilon)|},
\end{equation}
where $\vk_z(\epsilon) = \left[\vi-(\va^*+\epsilon
\vi)\right]^{-1}-\vi$.
We would like to let $\epsilon$ approach zero and consider the
limiting multiple description problem. In particular, we show that
\begin{eqnarray}
\vd_l(\epsilon) & \rightarrow & \vd_l, \quad \listl, \label{eq:deps1} \\
\vd_0(\epsilon) & \rightarrow & \vd_0^*, \label{eq:deps2}
\end{eqnarray}
as $\epsilon\rightarrow 0$ in Appendix~\ref{app:depsilon}. Further, we
show that
\begin{equation}
\vk_z(\epsilon)  \rightarrow (\vi-\va^*)^{-1}-\vi,\label{eq:eps3}
\end{equation}
as $\epsilon\rightarrow 0$ in Appendix~\ref{app:kzepsilon}.
Thus we can conclude that the sum rate approaches, using
\eqref{eq:enhanced0_1},
\begin{equation}
\frac{1}{2}\log\frac{|\vi+\vk_z|^{(L-1)}|\vd_0+\vk_z|}{|\vd_0|\prod\limits_{l=1}^L
|\vd_l+\vk_z|},
\end{equation}
as $\epsilon \rightarrow 0$; here $\vk_z = (\vi-\va^*)^{-1}-\vi$.
We observe that this sum rate is achievable using the jointly
Gaussian multiple scheme. Further, this sum rate is identical to
the lower bound to sum rate for the original distortions $(\vd_1,
\; \dots , \; \vd_L, \; \vd_0)$. Thus we conclude the optimality
of the jointly Gaussian description scheme in this case as well.

{\bf Case 3:} $\mathbf{0} \prec \va^* \preccurlyeq \vi$.
Alternatively, some eigenvalues of $\va^*$ are 1, but no
eigenvalues of $\va^*$ are 0. In this case, the $\vl_1 = {\bf 0}$
and the KKT conditions in \eqref{eq:kkt1}  reduce to
\begin{equation}
(\vk_{w_0}+\va^*)^{-1} -
\sum\limits_{l=1}^L(\vk_{\vw_l}+\va^*)^{-1}-\vl_2 = 0,
\end{equation}
for some $\vl_2  \succcurlyeq 0$ satisfying $\vl_2 (\va^*-\vi)  = {\bf
  0}$. Defining
$$
\vk_{w_l}^* = \left[
(\vk_{\vw_l}+\vi)^{-1}+\vl_2\right]^{-1}-\vi, $$
 we have, as in \eqref{eq:enhanced_1}, that
\begin{equation}\label{eq:enh1}
(\vk_{\vw_l}+\va^*)^{-1}+\vl_2  = (\vk_{\vw_l}^*+\va^*)^{-1}.
\end{equation}
The observation
\[
(\vk_{\vw_l}+\va^*)^{-1}+\vl_2 =
\left[(\vk_{\vw_l}+\vi)+(\va^*-\vi)\right]^{-1}+\vl_2,
\]
combined with the proof of \eqref{eq:enhanced_1} suffices to justify
\eqref{eq:enh1}. Now, from \eqref{eq:enh1},
\begin{equation}\label{eq:case3}
(\vk_{w_0}+\va^*)^{-1} -
\sum\limits_{l=1}^{L-1}(\vk_{w_l}+\va^*)^{-1}-(\vk_{w_L}^*+\va^*)^{-1}
= {\bf 0}.
\end{equation}
As in the previous case, the key step is to identify  smaller
distortion matrices at each of the individual receivers (ordered in the positive semidefinite sense) that is
achievable at the same sum rate:
$$\vd_l^* =\left[{\vk_{w_l}^*}^{-1}+\vi\right]^{-1}, \quad l = 1, \;
\ldots, \;
 L.$$
 To see that this is indeed
a smaller distortion matrix, observe that since $\vk_w$ is
positive definite, it follows that  $\vk_{w_l}^* \succ 0$ and
\begin{equation}
\begin{split}
\vd_l^* & =  \left[{\vk_{w_l}^*}^{-1}+\vi\right]^{-1} \\
& = \left[\left(\left(
(\vk_{w_l}+\vi)^{-1}+\vl_2\right)^{-1}-\vi\right)^{-1}+\vi\right]^{-1}
\\
& = \left[ \vi - (\vk_{w_l}+\vi)^{-1}-\vl_2\right] \\
& = \left[ \vi + \vk_{w_l} \right]^{-1} -\vl_2 \\
& = \vd_l - \vl_2, \quad l = 1, \; \ldots, \;
 L.
\end{split}
\end{equation}
Since $\vl_2 \succcurlyeq {\bf 0}$, it follows that   $\mathbf{0}
\prec \vd_l^* \preccurlyeq \vd_l, \; l=1, \; \ldots, \; L$. Define
\begin{equation}
\begin{split}
\vd_l(\epsilon) & = \left[(\vk_{w_l}+\epsilon
\vi)^{-1}+\vi\right]^{-1}, \quad l=0,\;1,\;\dots,\;L-1, \\
\vd_L(\epsilon) & = \left[(\vk_{w_L}^*+\epsilon
\vi)^{-1}+\vi\right]^{-1},
\end{split}
\end{equation}
then there exists $\delta >0$ such that for all $\epsilon \in
(0,\delta)$ we have $ \mathbf{0} \prec \va^*-\epsilon \vi \prec
\vi$, and $\mathbf{0} \prec \vd_l(\epsilon) \prec \vi$. We can
rewrite \eqref{eq:case3} as
\begin{equation}\label{eq:continues2}
\left[(\vk_{w_0}+\epsilon \vi)+(\va^*-\epsilon \vi)\right]^{-1} =
\sum\limits_{l=1}^{L-1}\left[(\vk_{w_l}+\epsilon
\vi)+(\va^*-\epsilon \vi)\right]^{-1}+\left[(\vk_{w_L}^*+\epsilon
\vi)+(\va^*-\epsilon \vi)\right]^{-1}.
\end{equation}
Thus if the distortion constraints were $(\vd_1(\epsilon), \;
\dots , \; \vd_L(\epsilon), \; \vd_0(\epsilon))$, then $
\va^*-\epsilon \vi$ is a solution to \eqref{eq:continues2}. This
situation corresponds to that discussed in Case I; we conclude
that the sum rate for this modified distortion multiple
description problem is
\begin{equation}
\frac{1}{2}\log\frac{|\vi+\vk_z(\epsilon)|^{(L-1)}|\vd_0(\epsilon)+\vk_z(\epsilon)|}{|\vd_0(\epsilon)|\prod\limits_{l=1}^L
|\vd_l(\epsilon)+\vk_z(\epsilon)|},
\end{equation}
where $\vk_z(\epsilon) = \left[\vi-(\va^*-\epsilon
\vi)\right]^{-1}-\vi$. We would like to let $\epsilon$ approach
zero and consider the limiting multiple description problem.
Similar to equations~\eqref{eq:deps1} and~\eqref{eq:deps2}, we
have
\begin{equation}
\begin{split}
\vd_l(\epsilon) & \rightarrow \vd_l,\quad l=1,\ldots L, \\
\vd_0(\epsilon) & \rightarrow \vd_0^*.
\end{split}
\end{equation}
Further, we show that
\begin{equation}\label{eq:kzinflemma}
\lim_{\epsilon \rightarrow
0}\frac{|\vi+\vk_z(\epsilon)|^{(L-1)}|\vd_0(\epsilon)+\vk_z(\epsilon)|}{|\prod\limits_{l=1}^L
|\vd_l(\epsilon)+\vk_z(\epsilon)|} = 1
\end{equation}
in Appendix \ref{app:kzinflemma}. We can now conclude that the
sum rate approaches
\begin{equation}
\frac{1}{2}\log\frac{1}{|\vd_0|}
\end{equation}
as $\epsilon$ approaches 0.
In other words, the point-to-point rate-distortion function for central
receiver with distortion $\vd_0$ can be achieved by using the jointly
Gaussian description scheme, and the resulting distortion is
$(\vd_1, \; \dots , \; \vd_L^*, \; \vd_0)$ where $\mathbf{0} \prec
\vd_L^* \preccurlyeq \vd_L$. In conclusion, the jointly Gaussian description
scheme is also optimal in this case.

{\bf Case 4:} $\mathbf{0} \preccurlyeq \va^* \preccurlyeq \vi$.
i.e., both 0 and 1 are eigenvalues of $\va^*$. In this case, the
KKT conditions are: there exist $\vl_1  \succcurlyeq 0$ and $
\vl_2 \succcurlyeq 0$ such that equations \eqref{eq:kkt1},
\eqref{eq:kkt2} and \eqref{eq:kkt3} hold. We can combine equations
\eqref{eq:enhanced_1} and \eqref{eq:enh1} to get
\begin{equation}\label{eq:case4}
(\vk_{w_0}^*+\va^*)^{-1} =
\sum\limits_{l=1}^{L-1}(\vk_{w_l}+\va^*)^{-1}+(\vk_{w_L}^*+\va^*)^{-1},
\end{equation}
where
\begin{equation*}
\begin{split}
\vk_{w_0}^* & = \left(\vk_{w_0}^{-1}+\vl_1\right)^{-1}, \\
\vk_{w_L}^* & = \left[ (\vk_{w_L}+\vi)^{-1}+\vl_2\right]^{-1}-\vi.
\end{split}
\end{equation*}

As in cases 2 and 3, we want to show the optimality of the jointly
Gaussian multiple
description scheme through a limiting procedure. We do this by first
perturbing $\va^*$ so that it has no eigenvalue equal to
0 or 1 as follows.

Without loss of generality, suppose that $\va^*$ has $p$ eigenvalues
equal to 0 and $q$ eigenvalues equal 1, where $p>0$ and $q>0$, and
there exists $N \times N$ orthogonal matrix $\vq$ such that
\[
\vq \va^* \vq^t =
\diag\{\underset{p}{\underbrace{0,\;\dots,\;0}},\underset{q}{\underbrace{\;1,\;\dots,\;1}},\;a_{p+q+1},\;\dots,\;a_N\},
\]
with $0 < a_{p+q+1} <1, \; \dots, 0 < a_{N} <1$. We need to
perturb the eigenvalues of $\va^*$ away from both 0 and 1. Towards this,
 we define two $N \times N$ diagonal matrices:
\begin{equation*}
\begin{split}
\mathbf{E}_1 & =
\diag(\underset{p}{\underbrace{1,\;\dots,\;1}},\underset{N-p}{\underbrace{\;0,\;\dots,\;0,\;0,\;\dots,\;0}}),
\\
\mathbf{E}_2 & =
\diag(\underset{p}{\underbrace{0,\;\dots,\;0}},\underset{q}{\underbrace{\;1,\;\dots,\;1}},\;0,\;\dots,\;0),
\end{split}
\end{equation*}
Also define
\begin{equation*}
\begin{split}
\va^*(\epsilon_1,\epsilon_2) & = \va^* + \vq^t (\epsilon_1
\mathbf{E}_1 -  \epsilon_2 \mathbf{E}_2)\vq, \\
\vk_z(\epsilon_1,\epsilon_2) & =
(\vi-\va^*(\epsilon_1,\epsilon_2))^{-1}-\vi, \\
\vk_{w_l}(\epsilon_1,\epsilon_2) & = \vk_{w_l} - \vq^t (\epsilon_1
\mathbf{E}_1 -  \epsilon_2 \mathbf{E}_2)\vq , \quad l=1,\;\dots,\;L-1,\\
\vk_{w_L}(\epsilon_1,\epsilon_2) & = \vk_{w_L}^* - \vq^t
(\epsilon_1
\mathbf{E}_1 -  \epsilon_2 \mathbf{E}_2)\vq, \\
\vk_{w_0}(\epsilon_1,\epsilon_2) & = \vk_{w_0}^* - \vq^t
(\epsilon_1 \mathbf{E}_1 -  \epsilon_2 \mathbf{E}_2)\vq.
\end{split}
\end{equation*}
Further,  defining
\begin{equation}
\vd_l(\epsilon_1,\epsilon_2) = (\vi +
\vk_{w_l}(\epsilon_1,\epsilon_2))^{-1}, \quad \listl,
\end{equation}
 there exists $\delta >0$ such that for all $\epsilon_1 \in
(0,\delta)$ and $\epsilon_2 \in (0,\delta)$ we have $ \mathbf{0}
\prec \va^*(\epsilon_1,\epsilon_2) \prec \vi$, and $\mathbf{0}
\prec \vd_l(\epsilon_1,\epsilon_2) \prec \vi$.
Now, we can rewrite \eqref{eq:case4} as
\begin{equation}\label{eq:continues3}
\Big[\vk_{w_0}(\epsilon_1,\epsilon_2)+\va^*(\epsilon_1,\epsilon_2)\Big]^{-1}
=
\sum\limits_{l=1}^{L}\Big[\vk_{w_l}(\epsilon_1,\epsilon_2)+\va^*(\epsilon_1,\epsilon_2)\Big]^{-1}.
\end{equation}
 Thus if the distortion
constraints were $(\vd_1(\epsilon_1, \epsilon_2), \; \dots , \;
\vd_L(\epsilon_1,\epsilon_2), \; \vd_0(\epsilon_1,\epsilon_2))$,
 then
$\va^*(\epsilon_1,\epsilon_2)$ is a solution to
\eqref{eq:continues3}. This situation corresponds to that
discussed in Case I; we conclude that the sum rate for this
modified distortion multiple description problem is
\begin{equation}
\frac{1}{2}\log\frac{|\vi+\vk_z(\epsilon_1,\epsilon_2)|^{(L-1)}|\vd_0(\epsilon_1,\epsilon_2)+
\vk_z(\epsilon_1,\epsilon_2)|}{|\vd_0(\epsilon_1,\epsilon_2)|\prod\limits_{l=1}^L
|\vd_l(\epsilon_1,\epsilon_2)+\vk_z(\epsilon_1,\epsilon_2)|},
\end{equation}
where $\vk_z(\epsilon_1,\epsilon_2) =
\left[\vi-\va^*(\epsilon_1,\epsilon_2)\right]^{-1}-\vi$. We would
like to let $\epsilon_1$ and $\epsilon_2$ approach zero and
consider the limiting multiple description problem. Similar to
equations~\eqref{eq:deps1} and~\eqref{eq:deps2}, when $\epsilon_1$
and $\epsilon_2$ approach 0, we get
\begin{equation}
\begin{split}
\vd_l(\epsilon_1,\epsilon_2) & \rightarrow \vd_l, \quad l=1,\;\dots,\;L-1, \\
\vd_L(\epsilon_1,\epsilon_2) & \rightarrow \vd_L^*, \\
\vd_0(\epsilon_1,\epsilon_2) & \rightarrow \vd_0^*,
\end{split}
\end{equation}
where $\vd_L^* = \vd_L-\vl_2$ as in case 3 and $\vd_0^* =
[\vd_0^{-1}+\vl_1^{-1}]^{-1}$ as in case 2. Further, we show that
\begin{equation}\label{eq:case4e1e2}
\lim_{\epsilon_2 \rightarrow 0}\lim_{\epsilon_1 \rightarrow 0}
\frac{1}{2}\log\frac{|\vi+\vk_z(\epsilon_1,\epsilon_2)|^{(L-1)}|\vd_0(\epsilon_1,\epsilon_2)+
\vk_z(\epsilon_1,\epsilon_2)|}{|\vd_0(\epsilon_1,\epsilon_2)|\prod\limits_{l=1}^L
|\vd_l(\epsilon_1,\epsilon_2)+\vk_z(\epsilon_1,\epsilon_2)|}
 = \frac{1}{2}\log \frac{1}{|\vd_0|}
\end{equation}
in Appendix \ref{app:case4e1e2}. We conclude that the sum
rate approaches
\begin{equation}
\frac{1}{2}\log\frac{1}{|\vd_0|}
\end{equation}
as $\epsilon_1$ and $\epsilon_2$ approach 0.
Thus the point-to-point rate-distortion function for central
receiver with distortion $\vd_0$ can be achieved by using the jointly
Gaussian description scheme, and the resulting distortions are
$(\vd_1, \; \dots , \; \vd_L^*, \; \vd_0^*)$ where $\mathbf{0}
\prec \vd_L^* \preccurlyeq \vd_L$ and $\mathbf{0} \prec \vd_0^*
\preccurlyeq \vd_0$. In other words, the  jointly Gaussian multiple description
scheme is also optimal in this case.

To summarize, we see that the jointly Gaussian description scheme
achieves the limiting sum rate. The limiting sum rate is the
solution to  an optimization problem. For some specific distortion
constraints,  the sum rate can be characterized as the solution to a
matrix polynomial equation (Case I).

\subsection{Rate Region for Two Descriptions}

Applying the result in Section \ref{sec:vecl} to the case of $L=2$,
i.e., the two description problem, we can see that jointly
Gaussian description scheme achieves the optimal sum rate. This
resolves the case left out in Section \ref{sec:vector2}. It also turns
out that in the two description problem, we can show that jointly
Gaussian description strategy achieves the entire rate region.
This is the main result of this subsection.

From Section \ref{sec:outerbound} we have a outer bound to the
rate region for the two description problem
\begin{equation}
\mathcal{R}_{out}(\vk_x,\;\vd_1,\;\vd_2,\;\vd_0) = \left\{
\begin{array}{l}
(R_1,\;R_2): \\
\displaystyle R_l \ge \frac{1}{2}\log\frac{|\vk_x|}{|\vd_l|}, \quad l=1,\;2 \\
\displaystyle R_1+R_2 \ge \sup_{\vk_z \succ \mathbf{0}}
\frac{1}{2}\log\frac{|\vk_x||\vk_x+\mathbf{K}_z||\mathbf{D}_0+\mathbf{K}_z|}
{|\mathbf{D}_0||\mathbf{D}_1+\mathbf{K}_z||\vd_2+\vk_z|}
\end{array}
\right\}.
\end{equation}


Following the discussion in Section \ref{sec:vecl}, we show in the
following that the jointly Gaussian description strategy (jointly
Gaussian multiple description schemes and the time sharing between
them) achieves the outer bound to the rate region.

Let
\[
\vk_{w_l} = (\vd_l^{-1} - \vk_x^{-1})^{-1}, \quad l=0, \; 1,\;2
\]
and
\[
F(\va) = \log|\vk_{w_0}+\va| -
\log|\vk_{w_1}+\va|-\log|\vk_{w_2}+\va|.
\]
Now consider the optimization problem:
\begin{equation}\label{eq:optimal_2}
\max\limits_{\mathbf{0} \preccurlyeq \va \preccurlyeq
\vk_x}\quad\quad F(\va).
\end{equation}
As in Section \ref{sec:vecl}, the optimal solution $\va^*$ falls
into four cases.

{\bf Case 1:}  $\mathbf{0} \prec \mathbf{A}^* \prec \vk_x$. In
this case, we know from Section \ref{sec:innerbound}  that the rate
pair $(R_1,R_2)$ satisfying
\begin{equation}
\left\{
\begin{array}{l}
(R_1,\;R_2): \\
\displaystyle R_l \ge \frac{1}{2}\log\frac{|\vk_x+\vk_{w_l}|}{|\vk_{w_l}|}, \quad l=1,\;2 \\
\displaystyle R_1+R_2 \ge
\frac{1}{2}\log\frac{|\vk_x+\vk_{w_1}||\vk_x+\vk_{w_2}|}{|\vk_{w}|}
\end{array}
\right\}
\end{equation}
is achievable using the jointly Gaussian multiple description scheme
with the covariance matrix of  $\vw_1, \; \vw_2$ being
\[
\vk_w = \begin{pmatrix}
                \vk_{w_1} & -\va^* \\
                -\va^* & \vk_{w_2}
       \end{pmatrix}.
\]
Denoting the resulting distortions as $(\vd_1, \; \vd_2, \; \vd_0)$,
we readily calculate
\[
 \frac{1}{2}\log\frac{|\vk_x+\vk_{w_l}|}{|\vk_{w_l}|} =
 \frac{1}{2}\log|
\vk_x||\vk_{w_l}^{-1}+\vk_x^{-1}| =
\frac{1}{2}\log\frac{|\vk_x|}{|\vd_l|}
\]
for $l=1,\;2$. From the discussion in Section \ref{sec:vecl},
we know that the lower bound  to sum rate is achieved using this
jointly Gaussian description scheme. Thus, in this case, the jointly
Gaussian description scheme achieves the rate region. As an aside, we
note  in
this case that, $\va^*$ satisfies
\[
\left[\vk^*_{w_0}+\va^*\right]^{-1} =
\left[\vk_{w_1}+\va^*\right]^{-1}+\left[\vk_{w_2}+\va^*\right],
\]
and,  from the discussion in Section \ref{sec:vector2}, that a
sufficient condition for this case to happen is \eqref{eq:vdfor2}.

{\bf Case 2:}  $\mathbf{0} \preccurlyeq \mathbf{A}^* \prec \vk_x$.
This case is similar to case 1: the jointly Gaussian description
scheme with covariance matrix for $\vw_1, \; \vw_2$ being
\[
\vk_w = \begin{pmatrix}
                \vk_{w_1} & -\va^* \\
                -\va^* & \vk_{w_2}
       \end{pmatrix}.
\]
achieves the lower bound on the rate region. We note that in this case the
resulting distortions are $(\vd_1, \; \vd_2, \; \vd_0^*)$, with
$\vd_0^* \preccurlyeq \vd_0$. Further,  we know from the discussion in
\ref{sec:vector2}, that a sufficient condition for this
case to happen is
\[
\vd_0^{-1} + \vk_x^{-1} - \vd_1^{-1}-\vd_2^{-1}  \preccurlyeq
\mathbf{0}.
\]

{\bf Case 3:}  $\mathbf{0} \prec \mathbf{A}^* \preccurlyeq \vk_x$.
In this case, we know from the discussion in Section \ref{sec:vecl}
 that for another two description problem with distortions
$(\vd_1,\;\vd_2^*,\;\vd_0)$ such that $\vd_2^* \preccurlyeq
\vd_2$, the jointly Gaussian description scheme with covariance
matrix for $\vw_1, \; \vw_2$ being
\[
\vk_w = \begin{pmatrix}
                \vk_{w_1} & -\va^* \\
                -\va^* & \vk^*_{w_2}
       \end{pmatrix}
\]
achieves the lower bound to sum rate
$\left(\frac{1}{2}\log\frac{|\vk_x|}{|\vd_0|}\right)$ to the
original distortions $(\vd_1,\;\vd_2,\;\vd_0)$. We can see, from the
contra-polymatroid structure of the achievable region of jointly
Gaussian description scheme,  that the corner point
$$B_1 =
\left(\frac{1}{2}\log\frac{|\vk_x|}{|\vd_1|},\frac{1}{2}\log\frac{|\vk_x|}{|\vd_0|}-\frac{1}{2}\log\frac{|\vk_x|}{|\vd_1|}\right)$$
in Figure \ref{fig:region} is achievable by this jointly Gaussian
description scheme.

Now observe that  the discussion in case 3 of Section
\ref{sec:vecl} is symmetric with respect to the  individual receivers.
Thus, by
exchanging the role of receiver 1 and receiver 2, we can achieve
the other corner point
$$B_2= \left(\frac{1}{2}\log\frac{|\vk_x|}{|\vd_0|}-\frac{1}{2}\log\frac{|\vk_x|}{|\vd_2|},\frac{1}{2}\log\frac{|\vk_x|}{|\vd_2|}\right)$$
in Figure \ref{fig:region} by another appropriate  jointly Gaussian description
scheme. Finally, time sharing between these two jointly Gaussian
multiple description schemes allows us to achieve the lower bound
on the rate region. As an aside, we note, as a consequence of  the
discussion in Section \ref{sec:vector2},  that a sufficient
condition for this case to happen is
\[
\vd_0+\vk_x - \vd_1 - \vd_2 \preccurlyeq \mathbf{0}.
\]

{\bf Case 4:}  $\mathbf{0} \preccurlyeq \mathbf{A}^* \preccurlyeq
\vk_x$. In this case, we know, from the discussion in Section
\ref{sec:vecl}, that for another two description problem
with distortions $(\vd_1,\;\vd_2^*,\;\vd^*_0)$ such that $\vd_2^*
\preccurlyeq \vd_2$ and $\vd_0^* \preccurlyeq \vd_0$, the jointly
Gaussian description scheme with covariance matrix for $\vw_1, \;
\vw_2$ being
\[
\vk_w = \begin{pmatrix}
                \vk_{w_1} & -\va^* \\
                -\va^* & \vk^*_{w_2}
       \end{pmatrix}
\]
achieves the lower bound to sum rate
$\left(\frac{1}{2}\log\frac{|\vk_x|}{|\vd_0|}\right)$ to the
original distortions $(\vd_1,\;\vd_2,\;\vd_0)$. Using an argument
entirely analogous to that applied  that the jointly Gaussian
description strategy achieves the rate region.

To summarize: the jointly Gaussian description strategy
achieves the rate region for the two description problem. For a class
of  distortion constraints, the corner points of the rate region can
be characterized by solving a matrix polynomial equation, as already
seen in Section \ref{sec:vector2}.

\section{Discussions}

Although multiple description for individual and central receivers
is a special case of the most general multiple description
problem, the solution to this problem  sheds substantial insight
to the issue-at-large. In this section, we discuss two instances
of other multiple description problems that can be resolved using
the insights developed so far. In particular, we discuss the
problem of two descriptions with separate distortion constraints
and the general multiple description problem for some special sets
of distortion constraints.

\subsection{Two Description with Separate Distortion Constraints}
\label{sec71} The problem of two descriptions with separate
distortion constraints is ilustrated in Figure
\ref{fig:md_special}.
Suppose the vector Gaussian  source $\vx[m]=(\vx_1[m],\vx_2[m])$,
the dimension of $\vx_1[m]$ is $N_1$ and the dimension of
$\vx_2[m]$ is $N_2$. This implies that the dimension of $\vx[m]$ is $N =
N_1+N_2$. Let $\vk_x = \meane[\vx[m]^t\vx[m]]$, $\vk_{x_1} =
\meane[\vx_1[m]^t\vx_1[m]]$, and $\vk_{x_2} =
\meane[\vx_2[m]^t\vx_2[m]]$. There are two encoders at the source
providing two descriptions of $\vx[m]$. There are three receivers:
the individual receivers 1 and 2 are only interested in generating
reproduction of $\vx_1[m]$ with mean square distortion constraint
$\vd_1$ (an $N_1 \times N_1$ positive definite matrix) from
description 1 and $\vx_2[m]$ with mean square distortion
constraint $\vd_2$ (an $N_2 \times N_2$ positive definite matrix)
from description 2, respectively. The central receiver uses both
descriptions to generate a reproduction of $\vx[m]$ with the error covariance
meeting a distortion constraint $\vd_0$ (an $N \times N$ positive definite
matrix) from both descriptions.

This situation is closely related to the two description problem and
we can harness our results thus far to completely characterize the
rate region of the problem at hand.
\begin{theorem}
The rate region of two description with separate distortion
constraints is
\begin{equation}\label{eq:innerboundseparate}
\mathcal{R}(\vd_1,\;\vd_2,\;\vd_0) =
\bigcup_{\Upsilon(\vd'_1,\;\vd'_2)}
\mathcal{R}_*(\vd'_1,\;\vd'_2,\;\vd_0),
\end{equation}
where $\Upsilon(\vd'_1,\;\vd'_2)$ is defined as
\begin{equation}
\Upsilon(\vd'_1,\;\vd'_2) \overset{\text{def}}{=} \left\{
\begin{array}{l}
(\vd'_1,\;\vd'_2): \; (\vd'_1)_{\{1, \dots,N_1\}}  \preccurlyeq
\vd_1, \; (\vd'_2)_{\{N_1+1, \dots,N\}}  \preccurlyeq \vd_2
\end{array}
\right\}.
\end{equation}
\end{theorem}
\begin{proof}
It is clear that any rate pair $(R_1,R_2) \in
\mathcal{R}_*(\vd'_1,\;\vd'_2,\;\vd_0)$ for some
$(\vd'_1,\;\vd'_2) \in \Upsilon(\vd'_1,\;\vd'_2)$ is in the rate
region for the two description with separate distortion
constraints, and so
\[ \mathcal{R}_*(\vd'_1,\;\vd'_2,\;\vd_0)
\subseteq \mathcal{R}(\vd_1,\;\vd_2,\;\vd_0).
\]
On the other hand, although receiver 1 (2) is only interested in
reconstructing $\vx_1$ ($\vx_2$), they can actually reconstruct the
entire source
$\vx$ based on their received descriptions. Hence, any coding
scheme for the two description with separate distortion
constraints will result in some achievable distortions
$(\vd'_1,\;\vd'_2,\;\vd'_0)$ with $(\vd'_1,\;\vd'_2) \in
\Upsilon(\vd'_1,\;\vd'_2)$ and $\vd_0' \preccurlyeq \vd_0$. Thus
any rate pair $(R_1,R_2) \in \mathcal{R}(\vd_1,\;\vd_2,\;\vd_0)$
achieved by this coding scheme is in the rate region
$\mathcal{R}_*(\vd'_1,\;\vd'_2,\;\vd_0)$ for the two description
problem. Thus
\[
\mathcal{R}(\vd_1,\;\vd_2,\;\vd_0) \subseteq
\bigcup_{\Upsilon(\vd'_1,\;\vd'_2)}
\mathcal{R}_*(\vd'_1,\;\vd'_2,\;\vd_0).
\]
From equivalence of the two regions in
\eqref{eq:innerboundseparate}, the proof is now complete.
\end{proof}

\subsection{General Gaussian Multiple Description Problem for
Special Choices of Distortion Constraints}
\label{sec72}
Consider the general Gaussian multiple description problem with
source covariance $\vk_x$ and $2^L-1$ distortion constraints $D_S$
for each $S \subseteq \{1, \; \dots, \; L\}$.

Following arguments similar to that used in arriving at  the lower bound
\eqref{eq:outer} for sum rate, we have an outer bound on the rate
region:
\begin{equation}\label{eq:generalregion}
\mathcal{R}_{out}(\vk_x, \; \vd_1,\;\dots,\;\vd_L,\;\vd_0) =
\left\{
\begin{array}{l}
(R_1,\;\dots,\;R_L): \\
\sum\limits_{l \in S}R_l \ge
\frac{1}{2}\log\frac{|\vk_x||\vk_x+\mathbf{K}_z|^{(|S|-1)}|\mathbf{D}_S+\mathbf{K}_z|}
{|\mathbf{D}_S|\prod\limits_{l \in S}|\mathbf{D}_l+\mathbf{K}_z|},
\quad \forall S \subseteq \{1, \; \dots, \; L\}
\end{array}
\right\}.
\end{equation}
Following  arguments similar to those used in arriving at  the
upper bound \eqref{eq:suminnerbound25} for the sum rate, we can use a
jointly Gaussian description scheme with covariance matrix of
$\vw_l$'s ($\vk_w$) taking the form \eqref{eq:kw},
any tuple $(R_1,\;\dots,\;R_L)$ satisfying
\begin{equation}
\left\{
\begin{array}{l}
(R_1,\;\dots,\;R_L): \\
\sum\limits_{l \in S}R_l \ge
\frac{1}{2}\log\frac{\Big|\vk_x\Big|\Big|\vk_x+\mathbf{K}_z\Big|^{(|S|-1)}\Big|\cov[\vx|\vu_l,\;l
\in S]+\mathbf{K}_z\Big|} {\Big|\cov[\vx|\vu_l,\;l \in
S]\Big|\prod\limits_{l \in
S}\Big|\cov[\vx|\vu_l]+\mathbf{K}_z\Big|},\quad \forall S
\subseteq \{1, \; \dots, \; L\}
\end{array}
\right\}
\end{equation}
is achievable. Thus if we can find a $\vk_w$ of the form in \eqref{eq:kw}
such that all of the $2^L-1$ distortion constraints are met with
equality, i.e.,
\begin{equation}
\vd_S = \cov[\vx|\vu_l,\;l \in S] = [\vk_x^{-1} + (\vi,\; \dots,
\; \vi)\mathbf{K}_{w_S}^{-1}(\vi,\; \dots, \; \vi)^t]^{-1} , \quad
\forall S \subseteq \{1, \; \dots, \; L\},
\end{equation}
where $\vk_{w_S}$ is the covariance matrix for all $\vk_{w_l}, l
\in S$, then the achievable region matches the outer bound and we
would have characterized the rate region of the multiple
description problem.

From the above discussion, we  see that for some choice of
distortion constraints of the multiple description problem, we can
indeed do this: First choose $L+1$ distortions
$(\vd_1,\;\vd_2,\;\dots,\;\vd_L,\;\vd_0)$ such that they satisfy
the condition for Theorem \ref{th:main} for the multiple
description problem with individual and central receivers. Next we
can solve  for the $\vk_w$ which is the covariance matrix of
$(\vw_1, \; \dots, \; \vw_L)$ for the sum-rate-achieving jointly
Gaussian description scheme. For any other $S \subseteq \{1, \;
\dots, \; L\}$, this scheme results in distortion $\vd_S =
[\vk_x^{-1} + (\vi,\; \dots, \; \vi)\mathbf{K}_{w_S}^{-1}(\vi,\;
\dots, \; \vi)^t]^{-1}$. Finally we choose these $\vd_S$'s as the
other distortion constraints. Now we have a general multiple
description problem with $2^L-1$ distortion constraints $D_S$ for
each $S \subseteq \{1, \; \dots, \; L\}$, and hence we can find a
$\vk_w$ of form \eqref{eq:kw} such that all of the $2^L-1$
distortion constraints are met with equality. Thus
\eqref{eq:generalregion} is actually the rate region and it can be
achieved by a jointly Gaussian description scheme.



\vspace{1cm}

\appendix
\noindent{\Large{\bf{Appendix}}}

\section{Useful Matrix Lemmas}\label{app:matrix}

In this appendix we provide some useful results in matrix analysis
that are extensively used in this paper.

\begin{lemma}[Matrix Inversion Lemma]\cite[Theorem 2.5]{Zhang99}
Let $\va$ be an $m \times m$ nonsingular matrix and $\mathbf{B}$
be an $n \times n$ nonsingular matrix and let $\mathbf{C}$ and
$\mathbf{D}$ be $m \times n$ and $n \times m$ matrices,
respectively. If the matrix $\va+\mathbf{CBD}$ is nonsingular,
then
\[
(\va+\mathbf{CBD})^{-1} =
\va^{-1}-\va^{-1}\mathbf{C}(\mathbf{B}^{-1}+
\mathbf{D}\mathbf{A}^{-1}\mathbf{C})^{-1}\mathbf{D}\va^{-1}
\]
\end{lemma}

\begin{lemma}\label{lemma:blockinverse}\cite[Theorem 2.3]{Zhang99}
Suppose that the partitioned matrix
\[
\mathbf{M} =
\begin{pmatrix}
\mathbf{A} & \mathbf{B} \\
\mathbf{C} & \mathbf{D}
\end{pmatrix}
\]
is invertible and that the inverse is conformally partitioned as
\[
\mathbf{M}^{-1} =
\begin{pmatrix}
\mathbf{X} & \mathbf{Y} \\
\mathbf{U} & \mathbf{V}
\end{pmatrix}.
\]
If $\va$ is a nonsingular principal sub-matrix of $\mathbf{M}$,
then
\begin{equation}
\begin{split}
\mathbf{X} = & \va^{-1} +
\va^{-1}\mathbf{B}(\vd-\mathbf{C}\va^{-1}\mathbf{B})^{-1}\mathbf{C}\va^{-1},
\\
\mathbf{Y} = &
-\va^{-1}\mathbf{B}(\vd-\mathbf{C}\va^{-1}\mathbf{B})^{-1}, \\
\mathbf{U} = & -
(\vd-\mathbf{C}\va^{-1}\mathbf{B})^{-1}\mathbf{C}\va^{-1}, \\
\mathbf{V} = & (\vd-\mathbf{C}\va^{-1}\mathbf{B})^{-1}.
\end{split}
\end{equation}
\end{lemma}

\begin{lemma}\cite[Theorem 6.13]{Zhang99}
Let $\mathbf{E} \in \mathbb{M}_n$ be a positive definite matrix
and let $\mathbf{F}$ be an $n \times m$ matrix. Then for any $m
\times m$ positive definite matrix $\mathbf{G}$,
\begin{equation}
\begin{pmatrix}
\mathbf{E} & \mathbf{F} \\
\mathbf{F}^t & \mathbf{G}
\end{pmatrix}
\succ 0  \Longleftrightarrow \mathbf{G} \succ
\mathbf{F}^t\mathbf{E}^{-1}\mathbf{F}.
\end{equation}
\end{lemma}

\begin{lemma}\cite[Theorem 6.8 and 6.9]{Zhang99}
Let $\va$ and $\mathbf{B}$ be positive definite matrices such that
$\va \succ \mathbf{B} \; (\va \succcurlyeq \mathbf{B})$. Then,
\begin{equation}
\begin{split}
 & |\va| \succ |\mathbf{B}| \quad ( |\va| \succcurlyeq
|\mathbf{B}| ),
\\
 & \va^{-1} \prec \mathbf{B}^{-1} \quad (  \va^{-1}
\preccurlyeq
\mathbf{B}^{-1}), \\
 & \va^{1/2} \succ \mathbf{B}^{1/2} \quad (\va^{1/2}
\succcurlyeq \mathbf{B}^{1/2}).
\end{split}
\end{equation}
\end{lemma}

\section{Proof of Lemma \ref{lemma:inform}}\label{app:inform}

Define an i.i.d.\ random process $\{\vz[m]\}$, $m=1,\; \dots,\; n$
of $\mathcal{N}(0, \mathbf{K}_z)$ Gaussian random vectors, where
$\vz[m]$, $m=1,\; \dots,\; n$ are independent of $\vx^n$ and
$C_l$, $l=1,\; \dots,\; L$. Form a random process
$\vy^n=(\vy[1],\;\dots,\;\vy[n])^t$ by
\[
\vy[m] = \vx[m]+\vz[m], \quad m=1,\; \dots,\; n.
\]
It follows that $\{\vy[m]\}$ is an i.i.d.\ random process
of $\mathcal{N}(0, \mathbf{K}_y)$ Gaussian random vectors, where
$\mathbf{K}_y = \vk_x + \mathbf{K}_z$. Then
\begin{equation}\label{eq:inform}
\begin{split}
I(C_1;\;& C_2; \;\dots;\;C_L)+I(C_1,\;\dots,\;C_L; \vx^n) \\
 = & \sum\limits_{l=1}^LH(C_l)-H(C_1,\cdots,C_L) +
I(C_1,\;\dots,\;C_L; \vx^n)  \\
 \ge & \sum\limits_{l=1}^LH(C_l)-H(C_1,\cdots,C_L) +
I(C_1,\;\dots,\;C_L; \vx^n) \\
&  - \Big(\sum\limits_{l=1}^LH(C_l|\vy^n)-H(C_1,\;\dots,\;C_L|\vy^n)\Big)  \\
 = &
\sum\limits_{l=1}^L\left(h(\vy^n)-h(\vy^n|C_l)\right)-h(\vy^n)
+h(\vy^n|C_1,\;\dots,\;C_L) + h(\vx^n)-h(\vx^n|C_1,\;\dots,\;C_L)  \\
 = & h(\vx^n)+(L-1)h(\vy^n)-\sum\limits_{l=1}^Lh(\vy^n|C_l)+
h(\vy^n|C_1,\;\dots,\;C_L)-h(\vx^n|C_1,\;\dots,\;C_L).
\end{split}
\end{equation}
Since $\vx^n$ and $\vy^n$ are Gaussian vectors, for the first two
terms in \eqref{eq:inform}, we have
\begin{equation}\label{eq:99}
\begin{split}
h(\vx^n) = & \frac{1}{2}\log(2\pi e)^{Nn}|\vk_x|^n, \\
h(\vy^n) =  & \frac{1}{2}\log(2\pi e)^{Nn}|\vk_y|^n =
\frac{1}{2}\log(2\pi e)^{Nn}|\vk_x+\vk_z|^n.
\end{split}
\end{equation}
We also have the following bound on $h(\vy^n|C_l)$ for
$l=1,\;\dots,\;L$:
\begin{equation}\label{eq:100}
\begin{split}
h(\vy^n|C_l) & \le \sum\limits_{m=1}^nh(\vy[m]|C_l) \\
& \le \sum\limits_{m=1}^n \frac{1}{2}\log(2\pi
e)^N\big|\cov[\vy[m]|C_l]\big| \\
& \le \frac{1}{2}\log(2\pi
e)^{Nn}+\frac{n}{2}\log\left|\frac{1}{n}\sum\limits_{m=1}^n
\cov[\vy[m]|C_l]\right| \\
& = \frac{1}{2}\log(2\pi
e)^{Nn}+\frac{n}{2}\log\left|\frac{1}{n}\sum\limits_{m=1}^n
\cov[(\vx[m]+\vz[m])|C_l]\right| \\
& = \frac{1}{2}\log(2\pi
e)^{Nn}+\frac{n}{2}\log\left|\frac{1}{n}\sum\limits_{m=1}^n
\cov[\vx[m]|C_l]+\vk_z\right| \\
& \le \frac{1}{2}\log(2\pi
e)^{Nn}+\frac{n}{2}\log\left|\vd_l+\vk_z\right| \\
& = \frac{1}{2}\log(2\pi e)^{Nn}\left|\vd_l+\vk_z\right|^n.
\end{split}
\end{equation}
Next we bound the last two terms of \eqref{eq:inform} as follows.
\begin{equation}\label{eq:12}
\begin{split}
 h(\vy^n|C_1,& \;\dots,\;C_L)-h(\vx^n|C_1,\;\dots,\;C_L)  \\
& =
h(\vy^n|C_1,\;\dots,\;C_L)-h(\vx^n|\vz^n,\;C_1,\;\dots,\;C_L)\\
& = h(\vy^n|C_1,\;\dots,\;C_L)-h(\vy^n|\vz^n,\;C_1,\;\dots,\;C_L)
\\
 & = I(\vy^n;\vz^n|C_1,\;\dots,\;C_L).
\end{split}
\end{equation}
Letting
\begin{equation}
\vk_c[m] \stackrel{\rm def}{=} \cov[\vx[m]-\hat{\vx}_0[m]],\label{eq:covdefn}
\end{equation}
we have
\begin{equation}\label{eq:awgnchannel}
\begin{split}
I(\vy^n;\vz^n|C_1,\;\dots,\;C_L) & =
h(\vz^n|C_1,\;\dots,\;C_L) - h(\vz^n|\vy^n,C_1,\;\dots,\;C_L) \\
& = h(\vz^n) - h(\vz^n|\vy^n-\hat{\vx}_0^n,C_1,\;\dots,\;C_L) \\
& \ge h(\vz^n) - h(\vz^n|\vy^n-\hat{\vx}_0^n) \\
& = \sum\limits_{m=1}^n
\big(h(\vz[m])-h(\vz[m]|\vz[1],\;\dots,\;\vz[m-1],\vy^n-\hat{\vx}_0^n)\big)
\\
& \ge \sum\limits_{m=1}^n
\big(h(\vz[m])-h(\vz[m]|\vy[m]-\hat{\vx}_0[m])\big) \\
& = \sum\limits_{m=1}^nI(\vz[m];\vx[m]-\hat{\vx}_0[m]+\vz[m]) \\
& \overset{(a)}{\ge}
\sum\limits_{m=1}^n\frac{1}{2}\log\frac{|\vk_c[m]+\vk_z[m]|}{|\vk_c[m]|}
\\ &
\overset{(b)}{\ge}\frac{n}{2}\log\frac{|\vd_0+\vk_z|}{|\vd_0|},
\end{split}
\end{equation}
where (a) is from \eqref{eq:covdefn} and \cite[Lemma II.2]{Diggavi01}.
The justfication for  (b) is from
the convexity of
$\log\frac{|\mathbf{A}+\mathbf{B}|}{|\mathbf{B}|}$ in $\mathbf{A}$
and \eqref{eq:lemmacov}. From \eqref{eq:12} and
\eqref{eq:awgnchannel} we have
\begin{equation}\label{eq:103}
h(\vy^n|C_1,\;\dots,\;C_L)-h(\vx^n|C_1,\;\dots,\;C_L) \ge
\frac{n}{2}\log\frac{|\vd_0+\vk_z|}{|\vd_0|}.
\end{equation}
Combining \eqref{eq:inform}, \eqref{eq:99} and \eqref{eq:103}, we
have
\begin{equation}\label{eq:104}
I(C_1;\; C_2; \;\dots;\;C_L)+I(C_1,\;\dots,\;C_L; \vx^n) \ge
\frac{n}{2}\log\frac{|\vk_x||\vk_x+\mathbf{K}_z|^{(L-1)}|\mathbf{D}_0+\mathbf{K}_z|}
{|\mathbf{D}_0|\prod\limits_{l=1}^L|\mathbf{D}_l+\mathbf{K}_z|}.
\end{equation}
Taking the supremum  over all positive definite
$\mathbf{K}_z$, we can sharpen the lower bound in \eqref{eq:104}:
\begin{equation}
\sum\limits_{l=1}^LI(C_1;C_2;\dots;C_L)+ I(C_1,\;\dots,\;C_L;
\vx^n) \ge \sup_{\vk_z \succ \mathbf{0}}
\frac{n}{2}\log\frac{|\vk_x||\vk_x+\mathbf{K}_z|^{(L-1)}|\mathbf{D}_0+\mathbf{K}_z|}
{|\mathbf{D}_0|\prod\limits_{l=1}^L|\mathbf{D}_l+\mathbf{K}_z|}.
\end{equation}

\section{Proof of Proposition~\ref{prop:nec_suff_condition}}
\label{app:nec_suff_condition}
Conditioned on $\vy$, the collection of random variables $(\vu_1,\;\dots, \; \vu_L)$ are
jointly Gaussian and thus we have
\begin{equation}
\sum\limits_{l=1}^Lh(\vu_l|\vy)-h(\vu_1,\;\dots,\;\vu_L|\vy) =
\frac{1}{2}\log\frac{\prod\limits_{l=1}^L
|\cov[\vu_l|\vy]|}{\big|\cov[\vu_1,\;\dots,\;\vu_L|\vy]\big|}.
\end{equation}
From MMSE of $\vu_l$ from $\vy$ we have
\begin{equation}
\cov[\vu_l|\vy]  =
\vk_x+\mathbf{K}_{w_l}-\vk_x(\vk_x+\mathbf{K}_z)^{-1}\vk_x, \quad
l=1,\;\dots,\; L
\end{equation}
and
\begin{equation}
  \cov(\vu_1,\;\dots,\;\vu_L|\vy)
    =
        \mathbf{J}\otimes \vk_x + \vk_w
        - \mathbf{J}\otimes \left(\vk_x(\vk_x+\mathbf{K}_z)^{-1}\vk_x\right),
\end{equation}
where $\mathbf{J}$ is an $L \times L $  matrix of all ones and
$\otimes$ is the Kronecker Product.

By Fischer inequality (the block matrix version of Hadamard
inequality, see \cite[Theorem 6.10]{Zhang99}) we know that
${\prod\limits_{l=1}^L |\cov[\vu_l|\vy]|} =
{\big|\cov[\vu_1,\;\dots,\;\vu_L|\vy]\big|}$ if and only if the
off-diagonal block matrices of $\cov[\vu_1,\;\dots,\;\vu_L|\vy]$
are all zero matrices. Thus we have
\begin{equation*}
\sum\limits_{l=1}^Lh(\vu_l|\vy)-h(\vu_1,\;\dots,\;\vu_L|\vy) = 0
\end{equation*}
if and only if
\begin{equation}\label{eq:kz}
\vk_x-\va = \vk_x(\vk_x+\vk_z)^{-1}\vk_x,
\end{equation}
or equivalently, if and only if
\begin{equation}
\mathbf{K}_z = \vk_x(\vk_x-\va)^{-1}\vk_x-\vk_x.
\end{equation}
To get a valid $\mathbf{K}_z \succ \mathbf{0}$, we need the
additional condition $ \mathbf{0} \prec \va \prec \vk_x$.

\section{Proof of Lemma \ref{lemma:inverse}}\label{app:inverse}

First we assume $\va \succ 0$, and hence
\begin{equation}
\begin{split}
 & \left[\va^{-1}+\left(\vi \; \vi \; \dots \; \vi\right)\vk_w^{-1}\left(\vi \; \vi \; \dots \;
\vi\right)^t \right]^{-1} \\
= & \va - \va \left(\vi \; \vi \; \dots
\vi\right)\left[\vk_w+\left(\vi \; \vi \; \dots \; \vi\right)^t
\va \left(\vi \; \vi
\; \dots \; \vi\right)\right]^{-1}\left(\vi \; \vi \; \dots \; \vi\right)^t \va \\
 = &  \va- \va \left(\vi \; \vi \; \dots \vi\right) \Big[\diag \{\vk_{w_1}+\va,\;
\vk_{w_2}+\va, \; \dots \; \vk_{w_L}+\va\} \Big]^{-1}\left(\vi \;
\vi \;
\dots \vi\right)^t \va \\
= & \va - \va \sum\limits_{l=1}^L [\vk_{w_l}+\va]^{-1} \va. \\
\end{split}
\end{equation}
Thus,
\begin{equation}
\begin{split}
(\vi \; \vi \; & \dots \; \vi)\vk_w^{-1}\left(\vi \; \vi \; \dots
\; \vi\right)^t \\
= & \left[\va - \va \sum\limits_{l=1}^L (\vk_{w_l}+\va)^{-1}
\va\right]^{-1}-\va^{-1} \\
= & \va^{-1} -  \va^{-1} \va \left[ -\left(\sum\limits_{l=1}^L
(\vk_{w_l}+\va)^{-1}\right)^{-1} + \va \va^{-1}\va\right]^{-1}\va\va^{-1} - \va^{-1} \\
= & \left[\left(\sum\limits_{l=1}^L
(\vk_{w_l}+\va)^{-1}\right)^{-1}-\va\right]^{-1}.
\end{split}
\end{equation}
When $\va$ is singular, we can choose $\delta >0$ such that
$\va+\epsilon \vi \succ 0$ for $\epsilon \in (0,\delta)$, and thus
we can apply the previous argument and let $\epsilon \rightarrow
0^+$ in the end.

\section{Proof of Lemma
\ref{lemma:positivedefinite}}\label{app:positivedefinite}

We use induction. First consider the  matrix
\[
\mathbf{\Delta}_2 = \begin{pmatrix}
\vk_{w_1} & -\va \\
-\va & \vk_{w_2}
\end{pmatrix}.
\]
We have
\begin{equation}
\begin{split}
\mathbf{\Delta}_2 \succ \mathbf{0} \; \Longleftrightarrow \;
&\vk_{w_2} \succ \va\vk_{w_1}^{-1}\va \\
\; \Longleftrightarrow \;
& \vk_{w_2}+\va \succ
\va\vk_{w_1}^{-1}\va+\va \\
\;  \Longleftrightarrow  \; & (\vk_{w_2}+\va)^{-1} \prec
(\va\vk_{w_1}^{-1}\va+\va)^{-1}  \\
\;  \Longleftrightarrow  \; &
(\vk_{w_2}+\va)^{-1} \prec
\va^{-1} - \left(\vk_{w_1}+\va\right)^{-1} \\
\;  \Longleftrightarrow  \; &
(\vk_{w_1}+\va)^{-1}+(\vk_{w_2}+\va)^{-1} \prec \va^{-1}
\\
\; {\Longleftarrow} \; &
\sum\limits_{l=1}^L\left(\vk_{w_l}+\va\right)^{-1} \prec
\va^{-1} \\
\;  \overset{(a)}{\Longleftrightarrow}  \; & (\vk_{w_0}+\va)^{-1}
\prec \va^{-1}
\\
\;  \Longleftrightarrow  \;  & \vk_{w_0}+\va \succ \va \\
\; \Longleftrightarrow  \; &  \vk_{w_0} \succ 0,
\end{split}
\end{equation}
where (a) is from \eqref{eq:core}.

Next we define
\[
\mathbf{\Delta}_k = \begin{pmatrix}
                \vk_{w_1} & -\va & -\va & \dots & -\va \\
                -\va & \vk_{w_2} & -\va & \dots & -\va \\
                \hdotsfor{5} \\
                -\va & \dots & -\va & \vk_{w_{k-1}} & -\va \\
                -\va & \dots & -\va & -\va & \vk_{w_k}
       \end{pmatrix}
\]
and suppose $\mathbf{\Delta}_k \succ 0$ for $k=3,\;\dots,\;l-1$.
Then
\begin{equation}
\begin{split}
\mathbf{\Delta}_l \succ 0 \;  \Longleftrightarrow  \; &  \vk_{w_l}
\succ
\va(\vi,\;\vi,\;\dots,\;\vi)\Delta_{l-1}^{-1}(\vi,\;\vi,\;\dots,\;\vi)^t\va \\
\;  \Longleftrightarrow  \; & \vk_{w_l} \succ
\va\left[\left(\sum\limits_{k=1}^{l-1}(\vk_{W_k}+\va)^{-1}\right)^{-1}-\va\right]^{-1}\va
\\
\;  \Longleftrightarrow  \; & \vk_{w_l}+\va \succ
\va\left[\left(\sum\limits_{k=1}^{l-1}(\vk_{w_k}+\va)^{-1}\right)^{-1}-\va\right]^{-1}\va
+\va \\
\;  \Longleftrightarrow  \; & (\vk_{w_l}+\va)^{-1} \prec
\left[\va\left[\left(\sum\limits_{k=1}^{l-1}(\vk_{w_k}+\va)^{-1}\right)^{-1}-\va\right]^{-1}\va
+\va\right]^{-1} \\
\;  \Longleftrightarrow  \; & (\vk_{w_l}+\va)^{-1} \prec \va^{-1}
-
\left[\left(\sum\limits_{k=1}^{l-1}(\vk_{w_k}+\va)^{-1}\right)^{-1}-\va+\va\right]^{-1}
\\
\;  \Longleftrightarrow  \; & (\vk_{w_l}+\va)^{-1} \prec \va^{-1}
-
\sum\limits_{k=1}^{l-1}(\vk_{w_k}+\va)^{-1} \\
\; {\Longleftarrow} \; & \sum\limits_{k=1}^{L}(\vk_{w_k}+\va)^{-1}
\prec
\va^{-1} \\
\;  \overset{(b)}{\Longleftrightarrow} \; & (\vk_{w_0}+\va)^{-1} \prec \va^{-1} \\
\;  \Longleftrightarrow  \; & \vk_{w_0}+\va \succ \va \\
\;  \Longleftrightarrow  \; & \vk_{w_0} \succ 0,
\end{split}
\end{equation}
where (b) is from \eqref{eq:core}.

\section{Proof of Proposition \ref{prop:twod}}\label{app:twod}
First we prove that
\[ \vd_0^{-1} + \vk_x^{-1} -
\vd_1^{-1}-\vd_2^{-1}  \succ \mathbf{0} \Rightarrow \va^* \succ
\mathbf{0}.
\]
\begin{proof}
We have
\begin{equation}
\va^* =
(\vk_{w_1}-\vk_{w_0})^{\frac{1}{2}}\left[(\vk_{w_1}-\vk_{w_0})^{-\frac{1}{2}}
(\vk_{w_2}-\vk_{w_0})(\vk_{w_1}-\vk_{w_0})^{-\frac{1}{2}}\right]^{\frac{1}{2}}(\vk_{w_1}-\vk_{w_0})^{\frac{1}{2}}
-\vk_{w_0}.
\end{equation}
Thus
\begin{equation}
\begin{split}
& \va^* \succ \mathbf{0} \\
\;  \Longleftrightarrow  \; &
(\vk_{w_1}-\vk_{w_0})^{\frac{1}{2}}\left[(\vk_{w_1}-\vk_{w_0})^{-\frac{1}{2}}
(\vk_{w_2}-\vk_{w_0})(\vk_{w_1}-\vk_{w_0})^{-\frac{1}{2}}\right]^{\frac{1}{2}}(\vk_{w_1}-\vk_{w_0})^{\frac{1}{2}}
\succ \vk_{w_0} \\
\;  \Longleftrightarrow  \;  &
\left[(\vk_{w_1}-\vk_{w_0})^{-\frac{1}{2}}
(\vk_{w_2}-\vk_{w_0})(\vk_{w_1}-\vk_{w_0})^{-\frac{1}{2}}\right]^{\frac{1}{2}}
\succ
(\vk_{w_1}-\vk_{w_0})^{-\frac{1}{2}}\vk_{w_0}(\vk_{w_1}-\vk_{w_0})^{-\frac{1}{2}}
\\
\; {\Longleftarrow} \; & (\vk_{w_1}-\vk_{w_0})^{-\frac{1}{2}}
(\vk_{w_2}-\vk_{w_0})(\vk_{w_1}-\vk_{w_0})^{-\frac{1}{2}}  \\
& \quad \succ
(\vk_{w_1}-\vk_{w_0})^{-\frac{1}{2}}\vk_{w_0}(\vk_{w_1}-\vk_{w_0})^{-1}\vk_{w_0}(\vk_{w_1}-\vk_{w_0})^{-\frac{1}{2}}
\\
\;  \Longleftrightarrow  \; & \vk_{w_2}-\vk_{w_0} \succ
\vk_{w_0}\left(\vk_{w_1}-\vk_{w_0}\right)^{-1}\vk_{w_0} \\
\;  \Longleftrightarrow  \; & \vk_{w_2}-\vk_{w_0} \succ
\vk_{w_0}\left(-\vi +
(\vk_{w_1}-\vk_{w_0})^{-1}\right)\vk_{w_1} \\
\;  \Longleftrightarrow  \; & \vk_{w_2}-\vk_{w_0} \succ -\vk_{w_0}
+
\vk_{w_0}(\vk_{w_1}-\vk_{w_0})^{-1}\vk_{w_1} \\
\;  \Longleftrightarrow  \; & \vk_{w_2} \succ
\vk_{w_0}(\vk_{w_1}-\vk_{w_0})^{-1}\vk_{w_1} \\
\;  \Longleftrightarrow  \; & \vk_{w_2} \succ
\vk_{w_0}\vk_{w_0}^{-1}(\vk_{w_0}^{-1}-\vk_{w_1}^{-1})^{-1}\vk_{w_1}^{-1}\vk_{w_1}
\\
\;  \Longleftrightarrow  \; & \vk_{w_2} \succ
(\vk_{w_0}^{-1}-\vk_{w_1}^{-1})^{-1} \\
\;  \Longleftrightarrow  \; & \vk_{w_1}^{-1} + \vk_{w_2}^{-1}
\prec
\vk_{w_0}^{-1} \\
\;  \Longleftrightarrow  \; & \vd_0^{-1} + \vk_x^{-1} -
\vd_1^{-1}-\vd_2^{-1} \succ \mathbf{0}.
\end{split}
\end{equation}
\end{proof}
The proof of
\[
\vd_0 + \vk_x - \vd_1 - \vd_2  \succ  \mathbf{0} \Rightarrow \va^*
\prec \vk_x
\]
is similar and hence is omitted.

\section{Proof of Lemma
\ref{lemma:enhanced0}}\label{app:enhanced0}

\begin{equation}
\begin{split}
\left[(\vk_{w_0}+\va^*)^{-1}+\vl_1\right]^{-1} & =
\left[(\vk_{w_0}+\va^*)^{-1}(\vi+
(\vk_{w_0}+\va^*)\vl_1)\right]^{-1} \\
& \overset{(a)}{=} (\vi+\vk_{w_0}\vl_1)^{-1}(\vk_{w_0}+\va^*) \\
& = (\vi+\vk_{w_0}\vl_1)^{-1}(\vk_{w_0}+\va^* -
(\vi+\vk_{w_0}\vl_1)\va^*)+\va^* \\
& \overset{(b)}{=} (\vi+\vk_{w_0}\vl_1)^{-1}\vk_{w_0}+\va^* \\
& = \left(\vk_{w_0}^{-1}(\vi+\vk_{w_0}\vl_1)\right)^{-1} +\va^* \\
& = \left(\vk_{w_0}^{-1}+\vl_1\right)^{-1} + \va^*,
\end{split}
\end{equation}
where (a) and (b) are from $\vl_1 \va^* = \mathbf{0}$.
\begin{equation}
\begin{split}
\frac{|\vd_0^*+\vk_z|}{|\vd_0^*|} & = |\vi+{\vd_0^*}^{-1}\vk_z| \\
& = |\vi+(\vd_0^{-1} + \vl_1)\vk_z| \\
& = |\vi+\vd_0^{-1}\vk_z + \vl_1\vk_z|\\
& = |\vi+\vd_0^{-1}\vk_z+\vl_1\left((\vi-\va^*)^{-1}-\vi\right)| \\
& \overset{(c)}{=} |\vi+\vd_0^{-1}\vk_z+\vl_1(\vi-\va^*)\left((\vi-\va^*)^{-1}-\vi\right)| \\
& = |\vi+\vd_0^{-1}\vk_z| \\
& = \frac{|\vd_0+\vk_z|}{|\vd_0|},
\end{split}
\end{equation}
where (c) is from $\vl_1 \va^* = \mathbf{0}$.

\section{Proof of Equations~\eqref{eq:deps1} and \eqref{eq:deps2}}\label{app:depsilon}

We first prove the following lemma.
\begin{lemma}\label{lemma:depsilon}
Let $\vd$ be an $N \times N$ matrix such that $\mathbf{0} \prec
\vd \prec \vi$. Let $\vk = (\vd^{-1}-\vi)^{-1}$. Choose $\epsilon
>0$ such that $\vk - \epsilon \vi \succ \mathbf{0}$. Define
\[
\vd(\epsilon) \overset{\text{def}}{=} \left[(\vk-\epsilon
\vi)^{-1}+\vi\right]^{-1}.
\]
Then, there exist constants $b_1 \ge b_2 >0$, such that
\begin{equation*}
\vd  - b_1 \epsilon \vi +o(\epsilon) \prec \vd(\epsilon) \prec \vd
- b_2 \epsilon \vi +o(\epsilon)
\end{equation*}
\end{lemma}
\begin{proof}
There exists  an $N \times N$ orthogonal matrix $\vq$ such that
\[ \vq \vk \vq^t = \diag\{k_1,\;\dots,\;k_N\}, \]
where $k_i>0$ are eigenvalues of $\vk$. We have
\begin{equation*}
\begin{split}
\vq \vd \vq^t & = \vq (\vk^{-1}+\vi)^{-1} \vq^t \\
& =
\diag\left\{\frac{k_1}{1+k_1},\;\dots,\;\frac{k_N}{1+k_N}\right\}, \\
\end{split}
\end{equation*}
and
\begin{equation*}
\begin{split}
\vq \vd(\epsilon) \vq^t & = \vq \Big[(\vk-\epsilon
\vi)^{-1}+\vi\Big]^{-1} \vq^t \\
& = \Big[(\diag\{k_1,\;\dots,\;k_N\}-\epsilon \vi)^{-1}+\vi\Big]^{-1} \\
& =
\diag\left\{\frac{k_1-\epsilon}{1+k_1-\epsilon},\;\dots,\;\frac{k_N-\epsilon}{1+k_N-\epsilon}\right\}\\
& =
\diag\left\{\frac{k_1}{1+k_1}-\frac{\epsilon}{(1+k_1)^2}+o(\epsilon
),\;\dots,\;\frac{k_N}{1+k_N}-\frac{\epsilon}{(1+k_N)^2}+o(\epsilon
)\right\}.
\end{split}
\end{equation*}
We now have
\begin{equation*}
\vq \vd \vq^t - b_1 \epsilon \vi +o(\epsilon) \prec \vq
\vd(\epsilon) \vq^t \prec \vq \vd \vq^t - b_2 \epsilon \vi
+o(\epsilon),
\end{equation*}
where $b_1 \ge b_2 > 0 $ are some constants. Hence
\begin{equation*}
\vd  - b_1 \epsilon \vi +o(\epsilon ) \prec \vd(\epsilon) \prec
\vd - b_2 \epsilon \vi +o(\epsilon ).
\end{equation*}
\end{proof}
Equations \eqref{eq:deps1} and \eqref{eq:deps2} are a direct
consequence of this lemma.

\section{Proof of Equation~\eqref{eq:eps3}}\label{app:kzepsilon}

We first prove the following lemma.
\begin{lemma}\label{lemma:kzepsilon}
Let $\va$ be an $N \times N$ matrix such that $\mathbf{0}
\preccurlyeq \va \prec \vi$. Let $\vk_z = (\vi-\va)^{-1}-\vi$.
Choose $\epsilon
>0$ such that $\va + \epsilon \vi \prec \vi$. Define
\[
\vk_z(\epsilon) \overset{\text{def}}{=} \left[\vi-(\va + \epsilon
\vi)\right]^{-1}-\vi.
\]
Then, there exist constants $c_1 \ge c_2 >0$ such that
\begin{equation*}
\vk_z  - c_1 \epsilon \vi +o(\epsilon ) \prec \vk_z(\epsilon)
 \prec  \vk_z  - c_2 \epsilon \vi +o(\epsilon ).
\end{equation*}
\end{lemma}
\begin{proof}
There exists an $N \times N$ orthogonal matrix $\vq$ such that
\[ \vq \va \vq^t = \diag\{a_1,\;\dots,\;a_N\} \]
where $a_i>0$ are the eigenvalues of $\va$. We have
\begin{equation*}
\begin{split}
\vq \vk_z \vq^t & = \vq ((\vi-\va)^{-1}-\vi) \vq^t \\
& =
\diag\left\{\frac{a_1}{1-a_1},\;\dots,\;\frac{a_N}{1-a_N}\right\}, \\
\end{split}
\end{equation*}
and
\begin{equation*}
\begin{split}
\vq \vk_z(\epsilon) \vq^t & = \vq ((\vi-(\va + \epsilon \vi))^{-1}-\vi) \vq^t \\
& =
\diag\left\{\frac{a_1+\epsilon}{1-a_1-\epsilon},\;\dots,\;\frac{a_N+\epsilon}{1-a_N-\epsilon}\right\}\\
& =
\diag\left\{\frac{a_1}{1-a_1}-\frac{(2a_1-1)\epsilon}{(1-a_1)^2}+o(\epsilon
),\;\dots,\;\frac{a_N}{1-a_N}-\frac{(2a_N-1)\epsilon}{(1-a_N)^2}+o(\epsilon
)\right\}.
\end{split}
\end{equation*}
We now  have
\begin{equation*}
\vq \vk_z \vq^t - c_1 \epsilon \vi +o(\epsilon) \prec \vq
\vk_z(\epsilon) \vq^t \prec \vq \vk_z \vq^t - c_2 \epsilon \vi
+o(\epsilon),
\end{equation*}
where $c_1 \ge c_2> 0 $ are some constants. Hence
\begin{equation*}
\vk_z  - c_1 \epsilon \vi +o(\epsilon ) \prec \vk_z(\epsilon)
 \prec \vk_z  - c_2 \epsilon \vi +o(\epsilon).
\end{equation*}
\end{proof}
Equation \eqref{eq:eps3} is a direct result of this lemma.

\section{Proof of equation \eqref{eq:kzinflemma}}\label{app:kzinflemma}

We first prove the following lemma.
\begin{lemma}\label{lemma:kzinf}
Let $\va$ be an $N \times N$ matrix such that $\mathbf{0} \prec
\va \preccurlyeq \vi$. Choose $\epsilon
>0$ such that $\va - \epsilon \vi \succ \vo$. Define
\[
\vk_z(\epsilon) \overset{\text{def}}{=} \left[\vi-(\va - \epsilon
\vi)\right]^{-1}-\vi.
\]
Then, for any $\mathbf{E}$ and $\mathbf{F}$ such that $\vo \prec
\mathbf{E} \preccurlyeq \vi$ and $\vo \prec \mathbf{F}
\preccurlyeq \vi$, we have
\begin{equation*}
\lim_{\epsilon \rightarrow
0}\frac{|\mathbf{E}+\vk_z(\epsilon)|}{|\mathbf{F}+\vk_z(\epsilon)|}
= 1.
\end{equation*}
\end{lemma}
\begin{proof}
There exists an $N \times N$ orthogonal matrix $\vq$ such that
\[ \vq \va \vq^t = \diag\{a_1,\;\dots,\;a_N\}, \]
where $0 < a_i \le 1$ are eigenvalues of $\va$. Without loss of
generality, we suppose $a_1=1, \;\dots,\; a_p=1, \; a_{p+1}<1,
\;\dots,\; a_N <1$.

We have
\begin{equation*}
\begin{split}
\vq \vk_z(\epsilon) \vq^t & = \vq ((\vi-(\va - \epsilon \vi))^{-1}-\vi) \vq^t \\
& = \diag\left\{\frac{1-\epsilon}{\epsilon},\;\dots,\;
\frac{1-\epsilon}{\epsilon}, \;
\frac{a_{p+1}-\epsilon}{1-a_{p+1}+\epsilon}, \;
\frac{a_N-\epsilon}{1-a_N+\epsilon}\right\},
\end{split}
\end{equation*}
and since
\begin{equation*}
\frac{|\mathbf{I}+\vk_z(\epsilon)|}{|\vk_z(\epsilon)|} \ge
\frac{|\mathbf{E}+\vk_z(\epsilon)|}{|\mathbf{F}+\vk_z(\epsilon)|}
\ge \frac{|\vk_z(\epsilon)|}{|\mathbf{I}+\vk_z(\epsilon)|},
\end{equation*}
we have
\begin{equation*}
\lim_{\epsilon \rightarrow
0}\frac{|\mathbf{E}+\vk_z(\epsilon)|}{|\mathbf{F}+\vk_z(\epsilon)|}
= 1.
\end{equation*}
\end{proof}
Equation \eqref{eq:kzinflemma} is a direct consequence of this lemma.

\section{Proof of Equation
\eqref{eq:case4e1e2}}\label{app:case4e1e2}

We would like to have a property similar to
\eqref{eq:enhanced0_1}, as $\epsilon_1$ approaches zero, and a
property similar to \eqref{eq:kzinflemma}, as $\epsilon_2$
approaches zero. To see this is the case, we need the following
lemma.
\begin{lemma}
\begin{equation*}
\vl_1 \vk_z(\epsilon_1=0,\epsilon_2) = \mathbf{0}
\end{equation*}
\end{lemma}
\begin{proof}
Since
\[ \vq \vl_1 \vq^t \vq \va^* \vq^t = \mathbf{0} \]
and
\begin{equation*}
\begin{split}
\vq \va^* \vq^t & =
\diag(\underset{p}{\underbrace{0,\;\dots,\;0}},\underset{q}{\underbrace{\;1,\;\dots,\;1}},\;a_{p+q+1},\;\dots,\;a_s)
\\
\vq \va^* \vq^t - \epsilon_2 \mathbf{E}_2 & =
\diag(\underset{p}{\underbrace{0,\;\dots,\;0}},\underset{q}{\underbrace{\;1-\epsilon_2,
\;\dots,\;1-\epsilon_2}},\;a_{p+q+1},\;\dots,\;a_s),
\end{split}
\end{equation*}
we have that
\[
\vq \va^* \vq^t (\vq \va^* \vq^t - \epsilon_2 \mathbf{E}_2) =
\mathbf{0}.
\]
Thus
\begin{equation*}
\begin{split}
\vq \vl_1 \vk_z(\epsilon_1=0,\epsilon_2)\vq^t & =  \vq \vl_1 \vq^t
\vq \left((\vi- \va^* + \vq^t
 \epsilon_2 \mathbf{E}_2 \vq )^{-1} - \vi\right) \vq^t  \\
& = \vq \vl_1 \vq^t \left((\vi - \vq \va^* \vq^t + \epsilon_2
\mathbf{E}_2)^{-1} -\vi\right) \\
& = \vq \vl_1 \vq^t (\vi - \vq \va^* \vq^t + \epsilon_2
\mathbf{E}_2) \left((\vi - \vq \va^* \vq^t + \epsilon_2
\mathbf{E}_2)^{-1} -\vi\right) \\
& = \mathbf{0}.
\end{split}
\end{equation*}
\end{proof}
Using this lemma, we can show a property similar to
\eqref{eq:enhanced0_1} as $\epsilon_1$ approaches zero. First note
that similar to case 2, we have
\begin{equation*}
\vd_0^{-1}+\vl_1  - e_2 \epsilon_2 \vi +o(\epsilon_2) \prec
\vd_0^{-1}(\epsilon_1, \epsilon_2) \prec \vd_0^{-1}+\vl_1 + e_1
\epsilon_1 \vi +o(\epsilon_1)
\end{equation*}
where $e_1 >0$ and $e_2 > 0$ are constants. Hence we have
\begin{equation*}
\begin{split}
\frac{|\vd_0(\epsilon_1=0, \epsilon_2)+\vk_z(\epsilon_1=0,
\epsilon_2)|}{|\vd_0(\epsilon_1=0, \epsilon_2)|} & =
|\vi+\vd^{-1}_0(\epsilon_1=0, \epsilon_2)\vk_z(\epsilon_1=0, \epsilon_2)| \\
& \ge |\vi+(\vd_0^{-1} + \vl_1 - e_2 \epsilon_2
\vi)\vk_z(\epsilon_1=0,
\epsilon_2)|\\
& = |\vi+\vd_0^{-1}\vk_z(\epsilon_1=0, \epsilon_2)-e_2
\epsilon_2\vk_z(\epsilon_1=0, \epsilon_2)| \\
& = \frac{|\vd_0+\vk_z(\epsilon_1=0, \epsilon_2)-e_2
\epsilon_2\vd_0\vk_z(\epsilon_1=0, \epsilon_2)|}{|\vd_0|}.
\end{split}
\end{equation*}
Similarly, we have
\begin{equation*}
\frac{|\vd_0(\epsilon_1=0, \epsilon_2)+\vk_z(\epsilon_1=0,
\epsilon_2)|}{|\vd_0(\epsilon_1=0, \epsilon_2)|} \le
\frac{|\vd_0+\vk_z(\epsilon_1=0, \epsilon_2)|}{|\vd_0|}.
\end{equation*}
Thus
\begin{equation}
\begin{split}
 \lim_{\epsilon_2  \rightarrow 0}\lim_{\epsilon_1 \rightarrow 0}
 & \frac{1}{2}\log\frac{|\vi+\vk_z(\epsilon_1,\epsilon_2)|^{(L-1)}|\vd_0(\epsilon_1,\epsilon_2)+
\vk_z(\epsilon_1,\epsilon_2)|}{|\vd_0(\epsilon_1,\epsilon_2)|\prod\limits_{l=1}^L
|\vd_l(\epsilon_1,\epsilon_2)+\vk_z(\epsilon_1,\epsilon_2)|} \\
& = \lim_{\epsilon_2 \rightarrow 0}
\frac{1}{2}\log\frac{|\vi+\vk_z(\epsilon_1=0,\epsilon_2)|^{(L-1)}|\vd_0+
\vk_z(\epsilon_1=0,\epsilon_2)|}{|\vd_0|\prod\limits_{l=1}^L
|\vd_l(\epsilon_1=0,\epsilon_2)+\vk_z(\epsilon_1=0,\epsilon_2)|} \\
& = \frac{1}{2}\log \frac{1}{|\vd_0|},
\end{split}
\end{equation}
where the last step is similar to \eqref{eq:kzinflemma}.

\bibliographystyle{IEEEtran}

\begin{thebibliography}{10}
\providecommand{\url}[1]{#1} \csname url@rmstyle\endcsname
\providecommand{\newblock}{\relax}
\providecommand{\bibinfo}[2]{#2}
\providecommand\BIBentrySTDinterwordspacing{\spaceskip=0pt\relax}
\providecommand\BIBentryALTinterwordstretchfactor{4}
\providecommand\BIBentryALTinterwordspacing{\spaceskip=\fontdimen2\font
plus \BIBentryALTinterwordstretchfactor\fontdimen3\font minus
  \fontdimen4\font\relax}
\providecommand\BIBforeignlanguage[2]{{%
\expandafter\ifx\csname l@#1\endcsname\relax
\typeout{** WARNING: IEEEtran.bst: No hyphenation pattern has been}%
\typeout{** loaded for the language `#1'. Using the pattern for}%
\typeout{** the default language instead.}%
\else \language=\csname l@#1\endcsname \fi #2}}

\bibitem{Ozarow80}
L.~Ozarow, ``On a source-coding problem with two channels and
three
  receivers,'' \emph{Bell Syst. Tech. J.}, vol.~59, no.~10, pp. 1909--1921,
  Dec. 1980.

\bibitem{ElGamal82}
A.~E. Gamal and T.~M. Cover, ``Achievable rates for multiple
descriptions,''
  \emph{IEEE Trans. Inform. Theory}, vol.~28, no.~6, pp. 851--857, Nov. 1982.

\bibitem{Ahlswede85}
R.~Ahlswede, ``The rate-distortion region for multiple
descriptions without
  excess rate,'' \emph{IEEE Trans. Inform. Theory}, vol.~31, no.~6, pp.
  721--726, Nov. 1985.

\bibitem{Zhang87}
Z.~Zhang and T.~Berger, ``New results in binary multiple
descriptions,''
  \emph{IEEE Trans. Inform. Theory}, vol.~33, no.~4, pp. 502--521, July 1987.

\bibitem{Zamir99}
R.~Zamir, ``Gaussian codes and shannon bounds for multiple
descriptions,''
  \emph{IEEE Trans. Inform. Theory}, vol.~45, no.~7, pp. 2629--2635, Nov. 1999.

\bibitem{FWFu02}
F.~W. Fu and R.~W. Yeung, ``On the rate-distortion region for
multiple
  descriptions,'' \emph{IEEE Trans. Inform. Theory}, vol.~48, no.~7, pp.
  2012--2021, July 2002.

\bibitem{Venkat03}
R.~Venkataramani, G.~Kramer, and V.~K. Goyal, ``Multiple
description coding
  with many channels,'' \emph{IEEE Trans. Inform. Theory}, vol.~49, no.~9, pp.
  2106--2114, Sept. 2003.

\bibitem{Pradhan04}
S.~S. Pradhan, R.~Puri, and K.~Ramchandran, ``n-channel symmetric
multiple
  descroption-part {I}: (n,k) source-channel erasure codes,'' \emph{IEEE Trans.
  Inform. Theory}, vol.~50, no.~1, pp. 47--61, Jan. 2004.

\bibitem{HYFeng05}
H.~Feng and M.~Effros, ``On the rate loss of multiple description
source
  codes,'' \emph{IEEE Trans. Inform. Theory}, vol.~51, no.~2, pp. 671--683,
  Feb. 2005.

\bibitem{Puri05}
R.~Puri, S.~S. Pradhan, and K.~Ramchandran, ``n-channel symmetric
multiple
  descroption-part {II}: an achievable rate-distortion region,'' \emph{IEEE
  Trans. Inform. Theory}, vol.~51, no.~4, pp. 1377--1392, Apr. 2005.

\bibitem{Vaishampayan93}
V.~A. Vaishampayan, ``Design of multiple description scalar
quantizers,''
  \emph{IEEE Trans. Inform. Theory}, vol.~39, no.~3, pp. 821--834, May 1993.

\bibitem{Vaishampayan94}
V.~A. Vaishampayan and J.~Domaszewicz, ``Design of
entropy-constrained
  multiple-description scalar quantizers,'' \emph{IEEE Trans. Inform. Theory},
  vol.~40, no.~1, pp. 245--250, Jan. 1994.

\bibitem{Vaishampayan98}
V.~A. Vaishampayan and J.~C. Batllo, ``Asymptotic analysis of
  multiple-description quantizers,'' \emph{IEEE Trans. Inform. Theory},
  vol.~44, no.~1, pp. 278--284, Jan. 1998.

\bibitem{Vaishampayan01}
V.~A. Vaishampayan, N.~Sloane, and S.~Servetto, ``Multiple
description vector
  quantizers with lattice codebooks: design and analysis,'' \emph{IEEE Trans.
  Inform. Theory}, vol.~47, no.~4, pp. 1718--1734, July 2001.

\bibitem{Goyal01}
V.~K. Goyal, ``Multiple description coding: compression meets the
network,''
  \emph{IEEE Signal Processing Mag.}, vol.~18, pp. 74--93, Sept. 2001.

\bibitem{Diggavi02}
S.~N. Diggavi, N.~J.~A. Sloane, and V.~A. Vaishampayan,
``Asymmetric multiple
  description lattice vector quantizers,'' \emph{IEEE Trans. Inform. Theory},
  vol.~48, no.~1, pp. 174--191, Jan. 2002.

\bibitem{Goyal02}
V.~K. Goyal, J.~A. Kelner, and J.~Kovacevic, ``Multiple
description vector
  quantization with a coarse lattice,'' \emph{IEEE Trans. Inform. Theory},
  vol.~48, no.~3, pp. 781--788, Mar. 2002.

\bibitem{CTian04}
C.~Tian and S.~S. Hemami, ``Universal multiple description scalar
quantizer:
  analysis and design,'' \emph{IEEE Trans. Inform. Theory}, vol.~50, no.~9, pp.
  2737--2751, Sept. 2004.

\bibitem{Ishwar03}
P.~Ishwar, R.~Puri, S.~S. Pradhan, and K.~Ramchandran, ``On
compression for
  robust estimation in sensor networks,'' in \emph{Proc. IEEE Int. Symp.
  Inform. Theory (ISIT)}, Yokohama, Japan, June-July 2003.

\bibitem{JChen05}
J.~Chen and T.~Berger, ``Robust distributed source coding,''
submitted to {\em
  IEEE Trans. Inform. Theory}, 2005.

\bibitem{Alon00}
N.~Alon and J.~H. Spencer, \emph{The probabilistic Method, 2nd
edition}.\hskip
  1em plus 0.5em minus 0.4em\relax New York: Wiley, 2000.

\bibitem{Welsh}
D.~J.~A.~Welsh, {\em Matroid Theory}, Academic Press, London,
1976.


\bibitem{Tse98}
D.~Tse and S.~Hanly, ``Multi-access fading channels: part {I}:
polymatroid
  structure,optimal resource allocation and throughput capacities,'' \emph{IEEE
  Trans. Inform. Theory}, vol.~44, no.~7, pp. 2796--2815, Nov. 1998.

\bibitem{Viswanath04}
P.~Viswanath, ``Sum rate of a class of gaussian multiterminal
source coding
  problems,'' in \emph{Advances in Network Information Theory}, P. Gupta, G.
  Kramer and A. Wijngaarden editors, Rutgers, NJ, 2004, pp. 43--60.

\bibitem{Zhang99}
F.~Zhang, \emph{Matrix Theory: Basic Results and
Techniques}.\hskip 1em plus
  0.5em minus 0.4em\relax Springer, 1999.

\bibitem{Diggavi01}
S.~Diggavi and T.~M. Cover, ``Worst additive noise under
covariance
  constraints,'' \emph{IEEE Trans. Inform. Theory}, vol.~47, no.~7, pp.
  3072--3081, Nov. 2001.


\end{thebibliography}


\end{document}